\documentclass{aa} 

\usepackage[varg]{txfonts}
\usepackage{natbib}
\usepackage{graphicx}
\usepackage{mathtools}
\usepackage{booktabs} 
\usepackage{multicol} 
\usepackage{multirow} 
\usepackage{array} 
\usepackage{graphicx}
\usepackage{mathabx} 
\usepackage{arydshln}
\usepackage{verbatim} 
\usepackage{subfigure} 
\usepackage[export]{adjustbox} 
\usepackage{color,epstopdf} 
	\DeclareGraphicsExtensions{.ps}
\begin{document}

\title{Exploring the volatile composition of comets C/2012 F6 (Lemmon) and C/2012 S1 (ISON) with ALMA}

\titlerunning{Volatiles in comets C/2012 F6 and C/2012 S1}


\author{Eva G. B\o gelund\inst{\ref{inst1}}
  \and Michiel R. Hogerheijde \inst{\ref{inst1}} 
    } 


\institute{Leiden Observatory, Leiden University, PO Box 9513, 2300
  RA, Leiden, The Netherlands\label{inst1}. \email{bogelund@strw.leidenuniv.nl}
	} 

\date{Submitted: 28/06/2016, Accepted: 04/05/2017}

\abstract
{Comets formed in the outer and cold parts of the disk which eventually evolved into our Solar System. Assuming that the comets have undergone no major  processing, studying their composition provides insight in the pristine composition of the Solar Nebula.}
{We derive production rates for a number of volatile coma species and explore how molecular line ratios can help constrain the uncertainties of these rates.} 
{We analyse observations obtained with the Atacama Large Millimetre/Submillimetre Array of the volatile composition of the comae of comets C/2012 F6 (Lemmon) and C/2012 S1 (ISON) at heliocentric distances of $\sim$1.45 AU and $\sim$0.56 AU, respectively. Assuming a Haser profile with constant outflow velocity, we model the line intensity of each transition using a 3D radiative transfer code and derive molecular production rates and parent scale lengths.}
{We report the first detection of CS in comet ISON obtained with the ALMA array and derive a parent scale length for CS of $\sim$200 km. Due to the high spatial resolution of ALMA, resulting in a synthesised beam with a size slightly smaller than the derived parent scale length (0$\overset{\second}{.}$59$\times$0$\overset{\second}{.}$39 corresponding to $\sim$(375$\times$250) km at the distance of the comet at the time of observations), we are able to tentatively identify CS as a daughter species, i.e., a species produced in the coma and/or sublimated from icy grains, rather than a parent species. In addition we report the detection of several CH$_3$OH transitions and confirm the previously reported detections of HCN, HNC and H$_2$CO as well as dust in the coma of each comet, and report 3$\sigma$ upper limits for HCO$^+$.}
{We derive molecular production rates relative to water of 0.2\% for CS, 0.06--0.1\% for HCN, 0.003--0.05\% for HNC, 0.1--0.2\% for H$_2$CO and 0.5--1.0\% for CH$_3$OH, and show that the modelling uncertainties due to unknown collision rates and kinematic temperatures are modest and can be mitigated by available observations of different transitions of HCN.}

 
\maketitle

\section{Introduction} \label{sec:introduction} 

Comets are generally believed to be leftover fragments of the protoplanetary disk that formed our solar system. Stored in the outer parts of the disk, these icy fragments are kept well away from the heat of the newborn Sun. While some comets may have been subject to subsequent processing through thermal heating and exposure to radiation when visiting the inner regions of the Solar System, others remain pristine. Therefore, the characterisation of cometary ices provides a unique opportunity to study the initial composition of the Solar Nebula.

In the classical picture comets are divided in two groups. The first group is comprised of the Jupiter-family comets. These were formed in the Kuiper Belt but now also populate the scattered disk. The second group is comprised of the long-period comets. These were formed in the region of the giant planets but now reside in the Oort Cloud \cite[see review by][]{Rickman2010}. Recent studies, though still debated, do not find this sharp division between groups of comets but suggest instead a much more extensive and continuous formation region around the CO and CO$_2$ snow lines \citep{AHearn2012}. In addition, the heterogeneity of the abundance of volatile species in comets indicates that a stationary formation scenario, where radial mixing is not accounted for, is unlikely \citep{Bockelee-Morvan2004}. On the other hand, a scenario in which comets are formed in a radially dynamic region of the disk fits well with the Grand Tack model \citep{Walsh2011}. In this model, inward and outward migrations of Jupiter and Saturn, during the first 100,000 years after the formation of the Sun, drove massive mixing in the disk.

In order to constrain the cometary formation sites further, it is essential to assess the composition of as many different comets as possible. To date, only a dozen comets have been studied in detail and only a handful of these in situ \cite[see, e.g,][for an overview]{AHearn2011}. The majority of studies show a vast compositional diversity amongst objects, demonstrating the need for more comprehensive statistical evaluations. Although in situ observations are undoubtedly the most precise and thorough way of quantifying cometary compositions, they are both expensive and rare. Therefore, we must rely largely on remote observations if we are to construct a statistically significant sample of objects, from which the emerging taxonomical database for comets can evolve \citep{Mumma2011}.

In this paper we analyse archival observations obtained with the Atacama Large Millimeter/Submillimeter Array (ALMA) of two comets. \citet{Cordiner2014, Cordiner2017b} present an analysis of some of this data; here we present detections of additional species, add analysis of the Band 6 Science Verification data, and check the consistency of our new analysis by comparison to these papers. Assuming a Haser model with constant outflow velocity we derive molecular production rates and parent scale lengths for each of the detected species. In addition, we explore how line ratios can be used to mitigate the uncertainty on the derived production rates due to the unknown kinetic temperature of each comet and the unknown collisional rates of water molecules with respect to other simple molecular species.

The paper is structured in the following way: in Section \ref{sec:observations} we summarise the observational setup, in Section \ref{sec:distribition_of_species} we present the observational results, in Section \ref{sec:model} we discuss our cometary model as well as the molecular production rates we derive and in Section \ref{sec:conclusion} we summarise our findings.

\section{Observations} \label{sec:observations}
C/2012 F6 (Lemmon) (hereafter referred to as Lemmon) is a long-period comet, with an orbital period $\sim$11,000 yr and semi-major axis $\sim$487AU, of high eccentricity, $e$=0.998, and orbital inclination, $i$=82.6$^{\degree}$. The comet was discovered on 2012 March 23 and reached perihelion one year later on 2013 March 24 at a distance of 0.73 AU\footnote{http://ssd.jpl.nasa.gov/?horizons; "JPL/HORIZONS 903922: COMET C/2012 F6 (LEMMON)". JPL Solar System Dynamics.}.

C/2012 S1 (ISON) (hereafter referred to as ISON) was a dynamically new, sungrazing comet, discovered on 2012 September 21. The comet reached perihelion on 2013 November 28 at a distance of merely 0.013 AU\footnote{"JPL/HORIZONS 903941: COMET C/2012 S1 (ISON)". JPL Solar System Dynamics.} ($\sim$2.7 R$_{\odot}$), after which it disintegrated \citep{Keane2016}.

\begin{table*}[t]
\begin{small}
  \centering
  \caption{Summary of Observations}
  \label{tab:obs_par}
    \begin{tabular}{ccccccccccc}
    \toprule
     Source & Setting & Species & Transition & Frequency & Date & Int. Time \tablefootmark{a} & r$_H$ \tablefootmark{b} & $\Delta$ \tablefootmark{c} & Ants. \tablefootmark{d} & Baselines \tablefootmark{e} \\
            & & & & (GHz) & & (min) & (AU) & (AU) & & (m) \\
    \midrule
		\multirow{8}{*}{\shortstack{C/2012 F6\\(Lemmon)}}
		& \multirow{4}{*}{I}
		& HCN           & 4-3                 & 354.50547 & 2013 May 31 - 2013 Jun 1  & 41.4 & 1.446 & 1.745 & 32 & 15-1284 \\
		& & H$_2$CO     & 5$_{1,5}$-4$_{1,4}$ & 351.76864 & & & & & & \\
		& & CH$_3$OH    & 7$_{2,5}$-6$_{2,4}$ & 338.72169 & & & & & & \\ 
		& &             & 7$_{1,6}$-6$_{1,5}$ & 341.41564 & & & & & & \\
		\cline{2-11} \\ [-2ex]
		& \multirow{2}{*}{II}
		& HNC           & 4-3                 & 362.63030 & 2013 May 30 - 2013 Jun 2  & 44.4 & 1.475 & 1.748 & 34 & 15-2733 \\  
		& & CH$_3$OH    & 1$_{1,1}$-0$_{0,0}$ & 350.90507 & & & & & & \\
		\cline{2-11} \\ [-2ex]
		& \multirow{3}{*}{SV}
		& HCN           & 3-2                 & 265.88640 & 2013 May 11               & 15.8 & 1.165 & 1.705  & 20 & 15-1175 \\
		& & HCO$^+$     & 3-2				  & 267.55753 & & & & & & \\
		& & CH$_3$OH    & 5$_{2,3}$-4$_{1,3}$ & 266.83812 & & & & & & \\
		\midrule		
		\multirow{5}{*}{\shortstack{C/2012 S1 \\(ISON)}}
		& \multirow{3}{*}{I}
		& CS            & 7-6                 & 342.88286 & 2013 Nov 16 - 2013 Nov 17 & 68.5 & 0.589 & 0.887 & 29 & 17-1284 \\ 
		& & HCN         & 4-3                 & 354.50547 & & & & & & \\ 
		& & HCO$^+$     & 4-3                 & 356.73424 & & & & & & \\ 
		\cline{2-11} \\ [-2ex]
		& \multirow{3}{*}{II}
        & HNC           & 4-3                 & 362.63030 & 2013 Nov 16 - 2013 Nov 17 & 91.7 & 0.557 & 0.875 &  29 & 12-1284 \\
        & & H$_2$CO     & 5$_{1,5}$-4$_{1,4}$ & 351.76864 & & & & & & \\ 
        & & CH$_3$OH    & 1$_{1,1}$-0$_{0,0}$ & 350.90507 & & & & & & \\  
    \bottomrule
    \end{tabular}
    \tablefoot{\tablefoottext{a}{Total time on source}. \tablefoottext{b}{Heliocentric distance}. \tablefoottext{c}{Geocentric distance}. \tablefoottext{d}{Number of 12m antennas in array}. \tablefoottext{e}{Deprojected antenna baseline range}}
\end{small}	
\end{table*}

Observations of both comets were carried out with ALMA in Cycle 1 Early Science mode between 30 May and 2 June 2013 and 16 and 17 November 2013 for comets Lemmon and ISON respectively, using the ALMA Band 7 receivers (covering the frequency range 275-373 GHz). 

Simultaneous observations of two sets of spectral lines (plus continuum) were made for each comet with the correlators configured to cover four frequency ranges in each set. Table \ref{tab:obs_par} summarises these and general observation parameters (for clarity, spectral windows with no detections are not listed). All spectral windows have a total bandwidth of 937500 kHz with 3840 equally spaced channels providing a spectral resolution of 244 kHz corresponding to $\sim$0.20 km s$^{-1}$ at 373 GHz. The cometary positions were traced using JPL Horizons ephemerides (JPL$\#$78 for Lemmon and JPL$\#$54 for ISON).

The quasars  3C279, J0006-0623, J2232+117 and J0029+3456 were used for phase and bandpass calibration while flux scale calibrations were done for Lemmon using Pallas and for ISON using Ceres and Titan. Weather conditions were good with low precipitable water vapour (PWV) at zenith between 0.44-0.83 mm and high atmospheric phase stability.

Additional observations of comet Lemon were made on 2013 May 11 as part of the Science Verification (SV) program to test the capability of the array to Doppler track ephemeris targets. These observations were carried out using the Band 6 receivers (covering the frequency range 211-275 GHz), targeting four spectral lines (see Table \ref{tab:obs_par}). The SV data bands have total widths of 234375 kHz with 3840 channels. The resolution is 61 kHz corresponding to 0.07 km s$^{-1}$ at 275 GHz. Phase, bandpass and absolute flux scale calibrations are done using the quasars J2232+117, J0006-0623 and J0238+166. 

We optimise the standard data delivery reduction scripts for each target and use these to flag and calibrate the data. Deconvolution is done in CASA 4.2.2 using the CLEAN algorithm with natural visibility weighting. The image pixel size was set to 0$\overset{\second}{.}$1$\times$0$\overset{\second}{.}$1. This resolution element corresponds to (127$\times$127) km and (64$\times$64) km at the distances of Lemmon and ISON respectively, on the days of observations. The images were restored with a Gaussian beam between 0$\overset{\second}{.}$79$\times$0$\overset{\second}{.}$50 and 0$\overset{\second}{.}$55$\times$0$\overset{\second}{.}$37 (depending on the line frequency of the individual transitions) and the spectral coordinates were shifted to the rest-frequency of the targeted lines. As noted by \cite{Cordiner2014}, no signal was detected on baselines longer than $\sim$500 m so we excluded these during imaging to avoid introducing unnecessary noise. This reduces the angular resolution of the data slightly but all species are still well-sampled. 

For each set of observations we image both the continuum and individual line emission. We assume the cometary nucleus to be at the position of the continuum peak, which can be clearly identified in all images. However, due to the nature of comets as non-gravitationally accelerating bodies, these are offset from the arrays pointing centre by $\sim$0$\overset{\second}{.}$9 in the case of Lemmon and $\sim$6$\overset{\second}{.}$5 in the case of ISON. To account for the offset of the pointing centre with respect to the position of the cometary nucleus, images were primary beam corrected. One execution block of observations of ISON was excluded completely due to incorrect tracking.  

\section{Spatial distribution of molecules}
\label{sec:distribition_of_species}

\begin{figure*}
	\centering
	\subfigure{\label{} 
		\includegraphics[width=0.248\textwidth, trim={0.3cm 1.4cm 0 0.4cm}, clip]{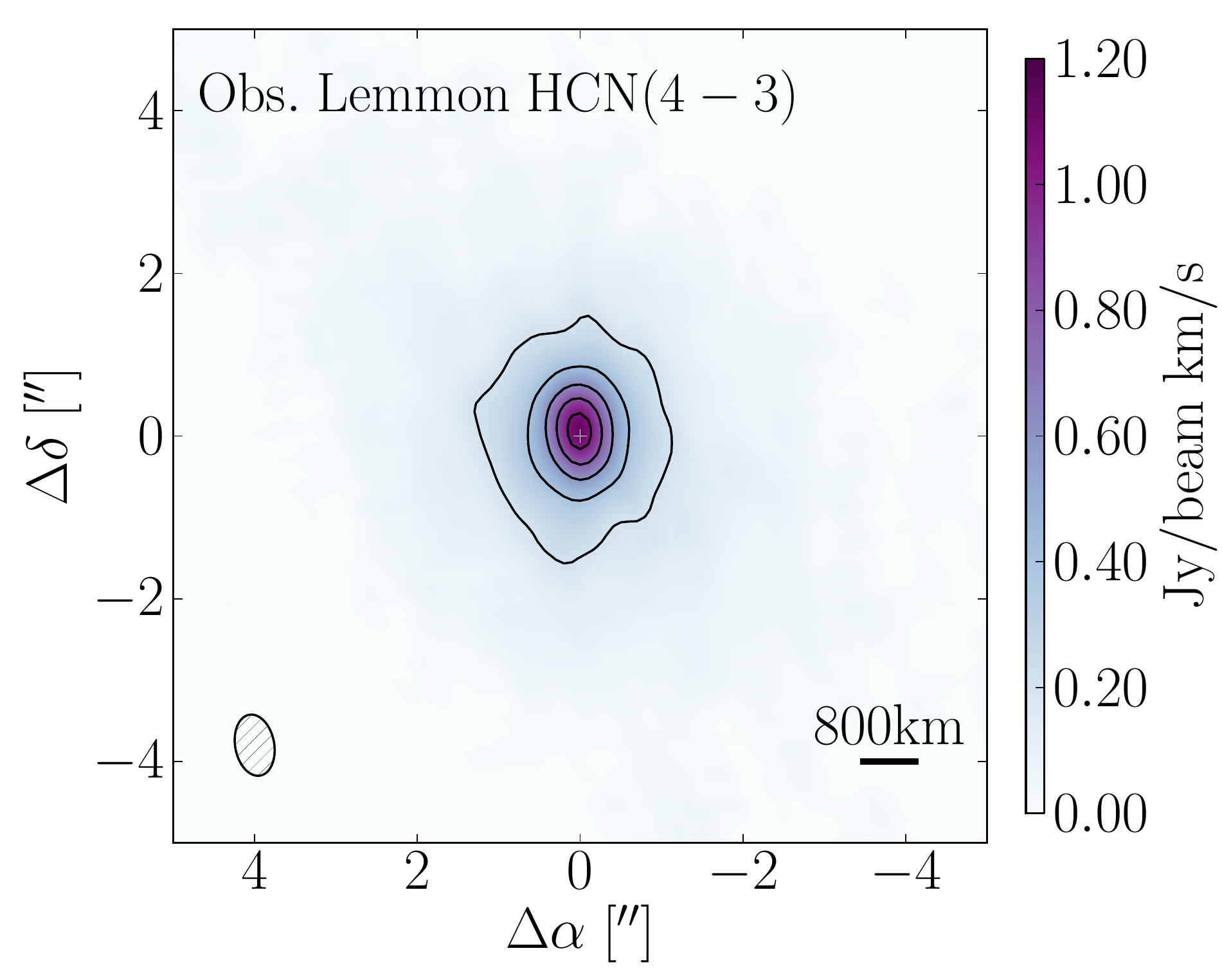}} 
	\subfigure{\label{}
		\includegraphics[width=0.233\textwidth, trim={1.4cm 1.4cm 0 0.1cm}, clip]{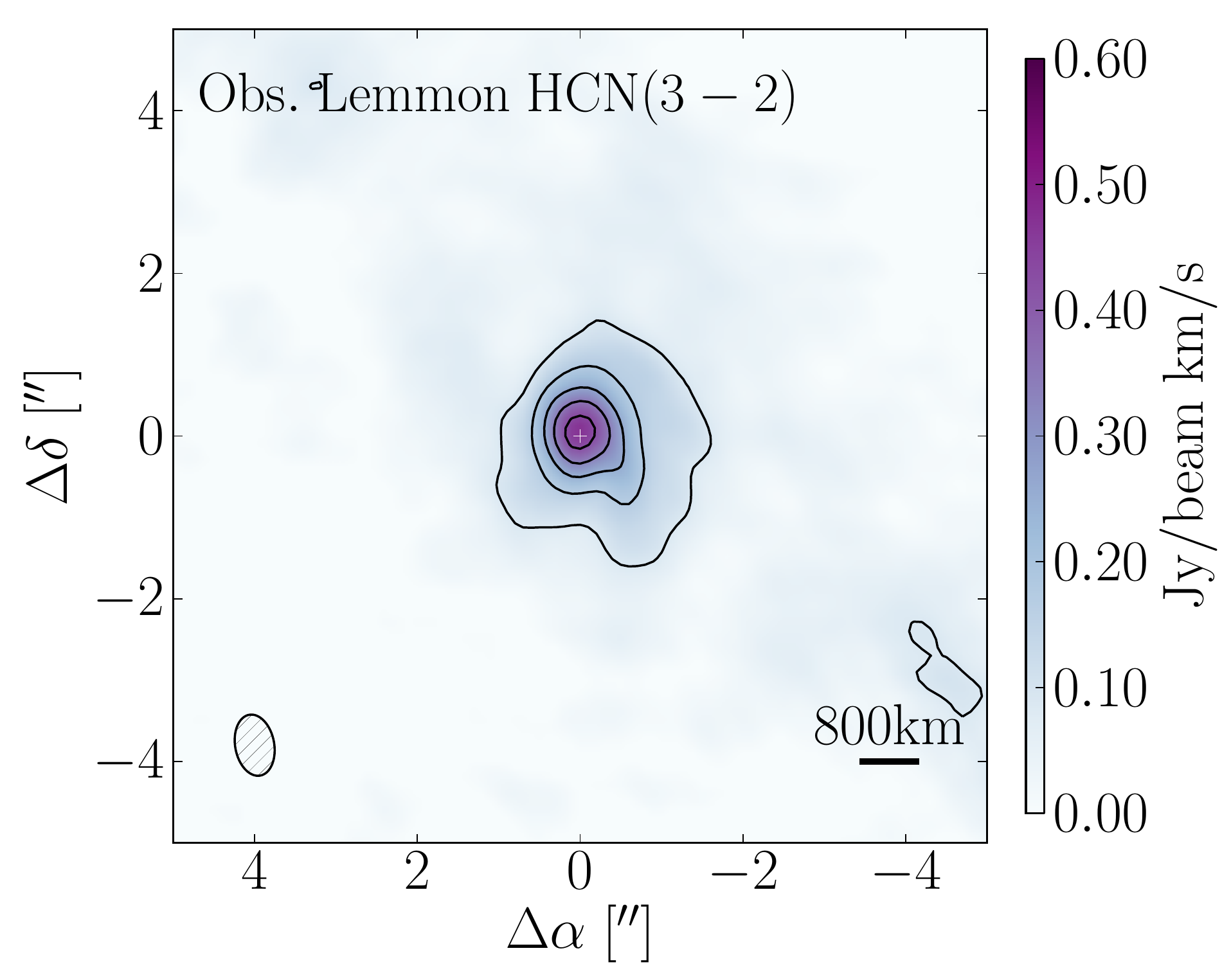}}  
	\subfigure{\label{} 
		\includegraphics[width=0.233\textwidth, trim={1.4cm 1.4cm 0 0.1cm}, clip]{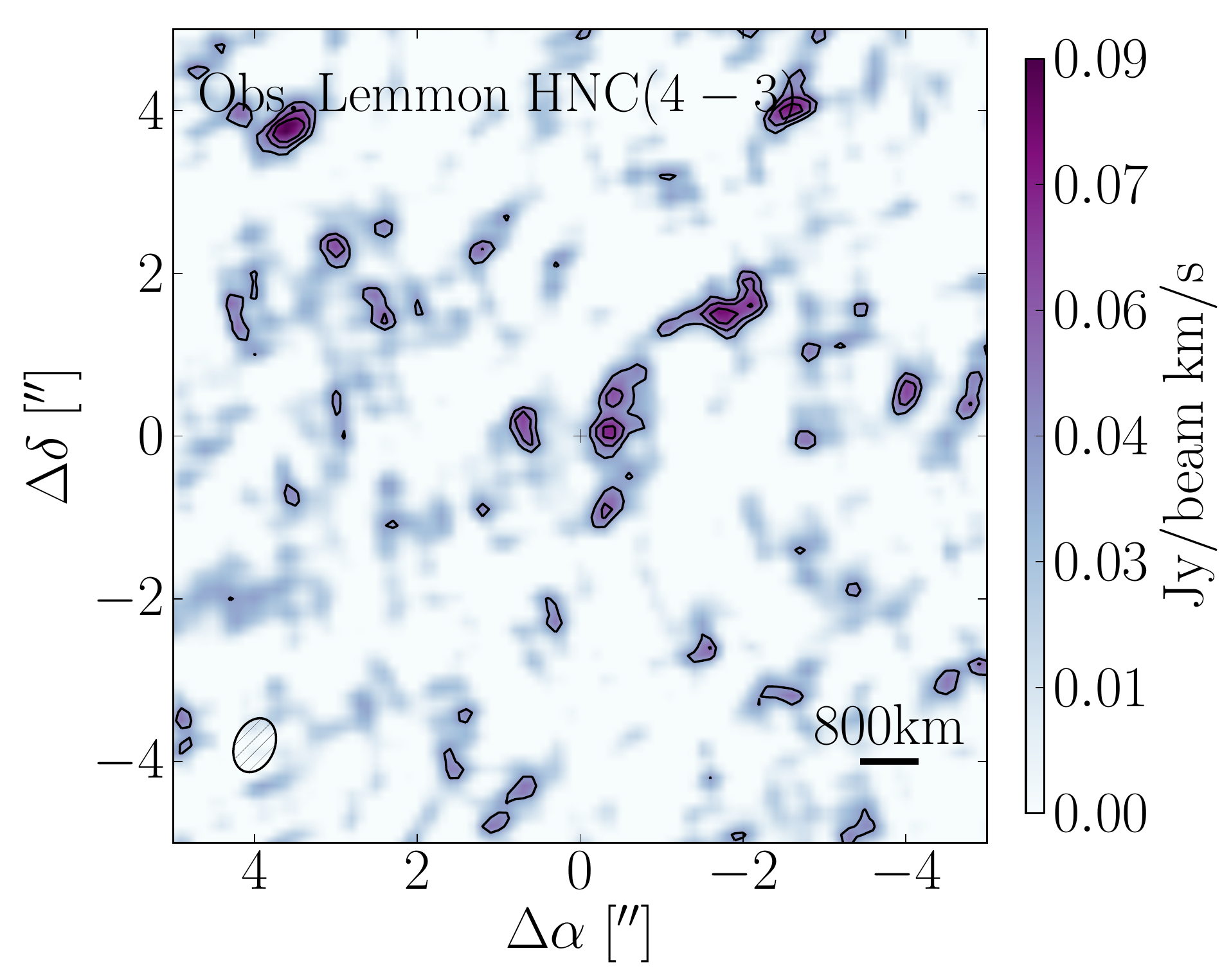}}
	\subfigure{\label{}
		\includegraphics[width=0.233\textwidth, trim={1.4cm 1.4cm 0 0.1cm}, clip]{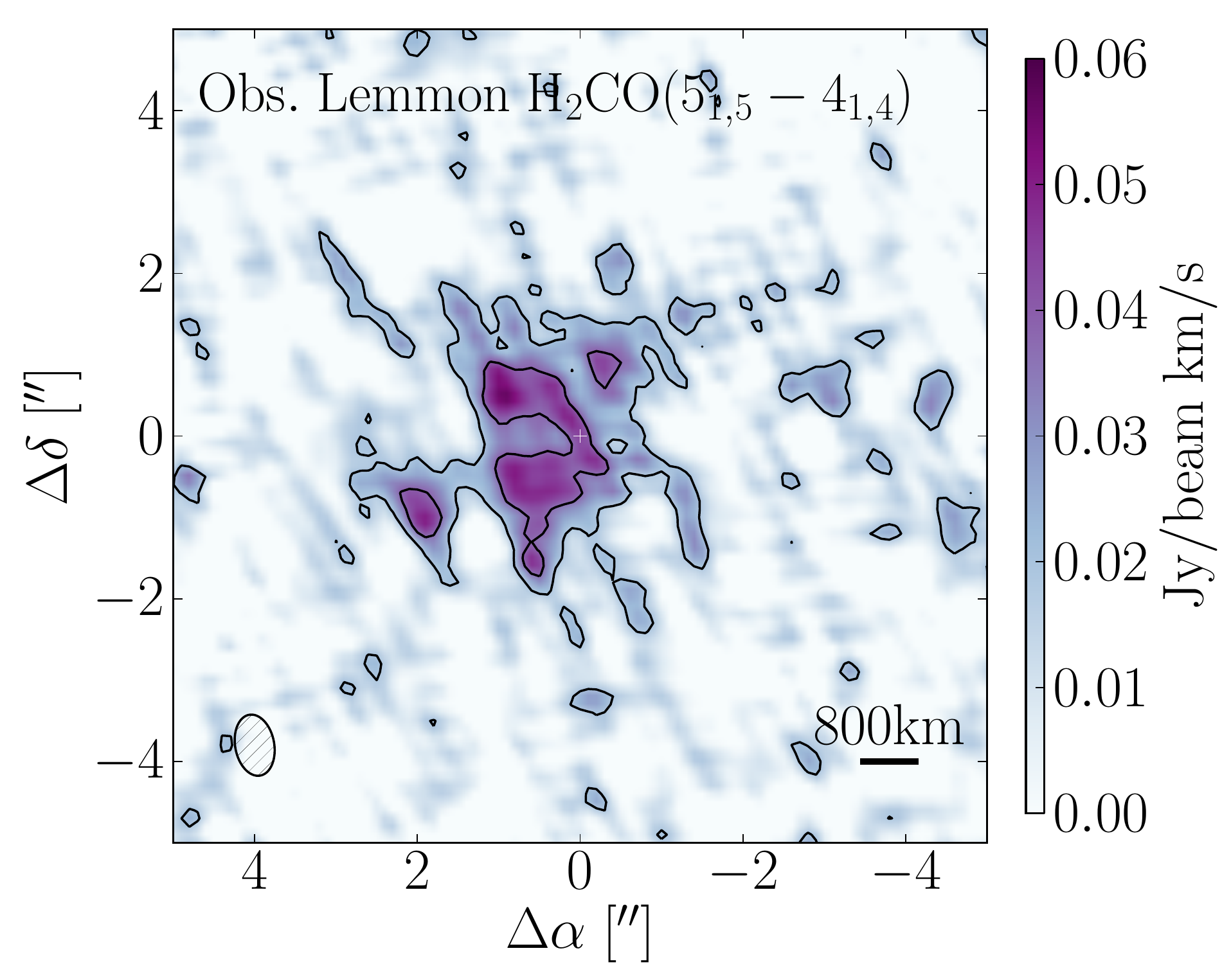}}
	
	\subfigure{\label{} 
		\includegraphics[width=0.248\textwidth, trim={0.3cm 1.4cm 0 0.4cm}, clip]{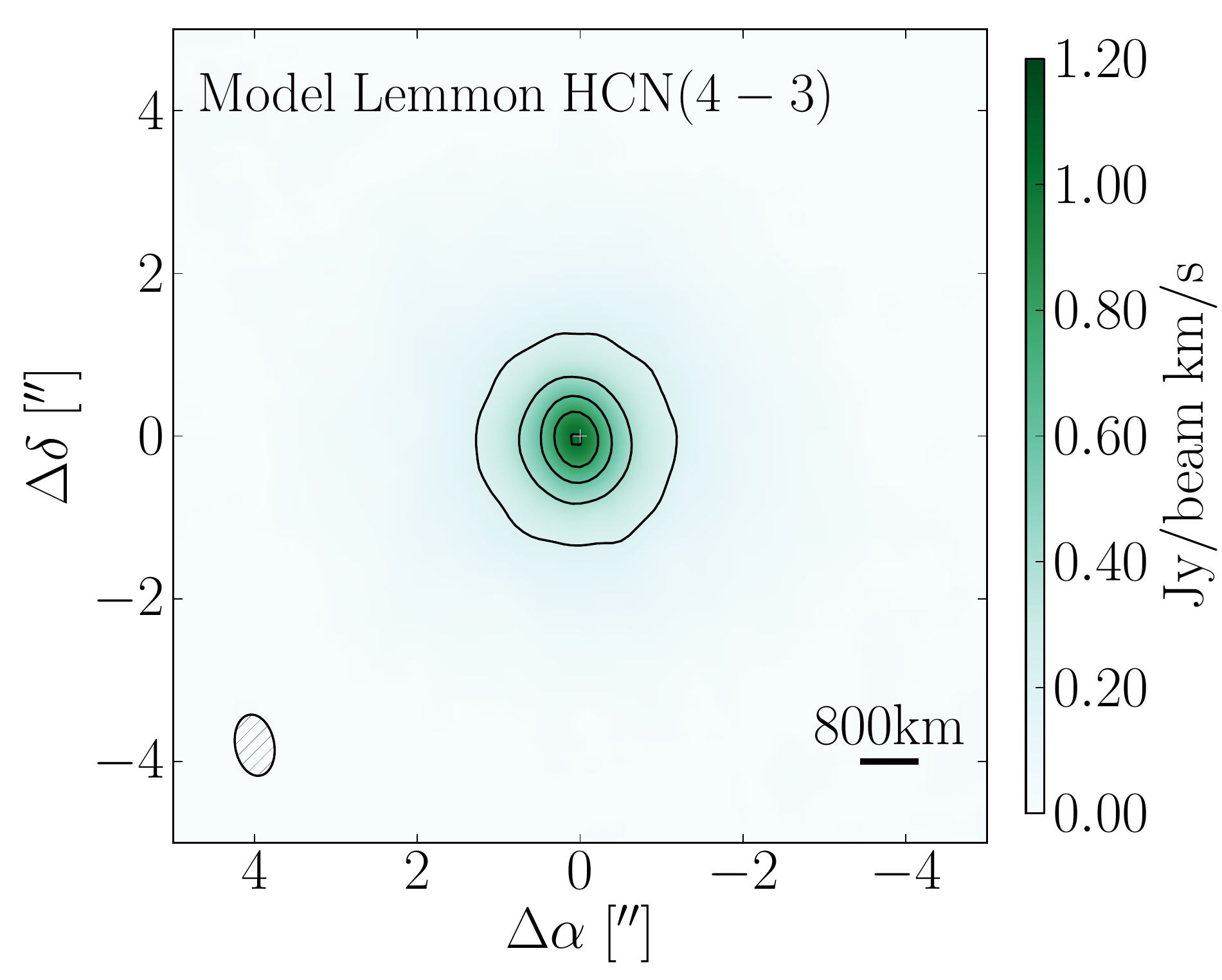}} 
	\subfigure{\label{}
		\includegraphics[width=0.233\textwidth, trim={1.4cm 1.4cm 0 0.1cm}, clip]{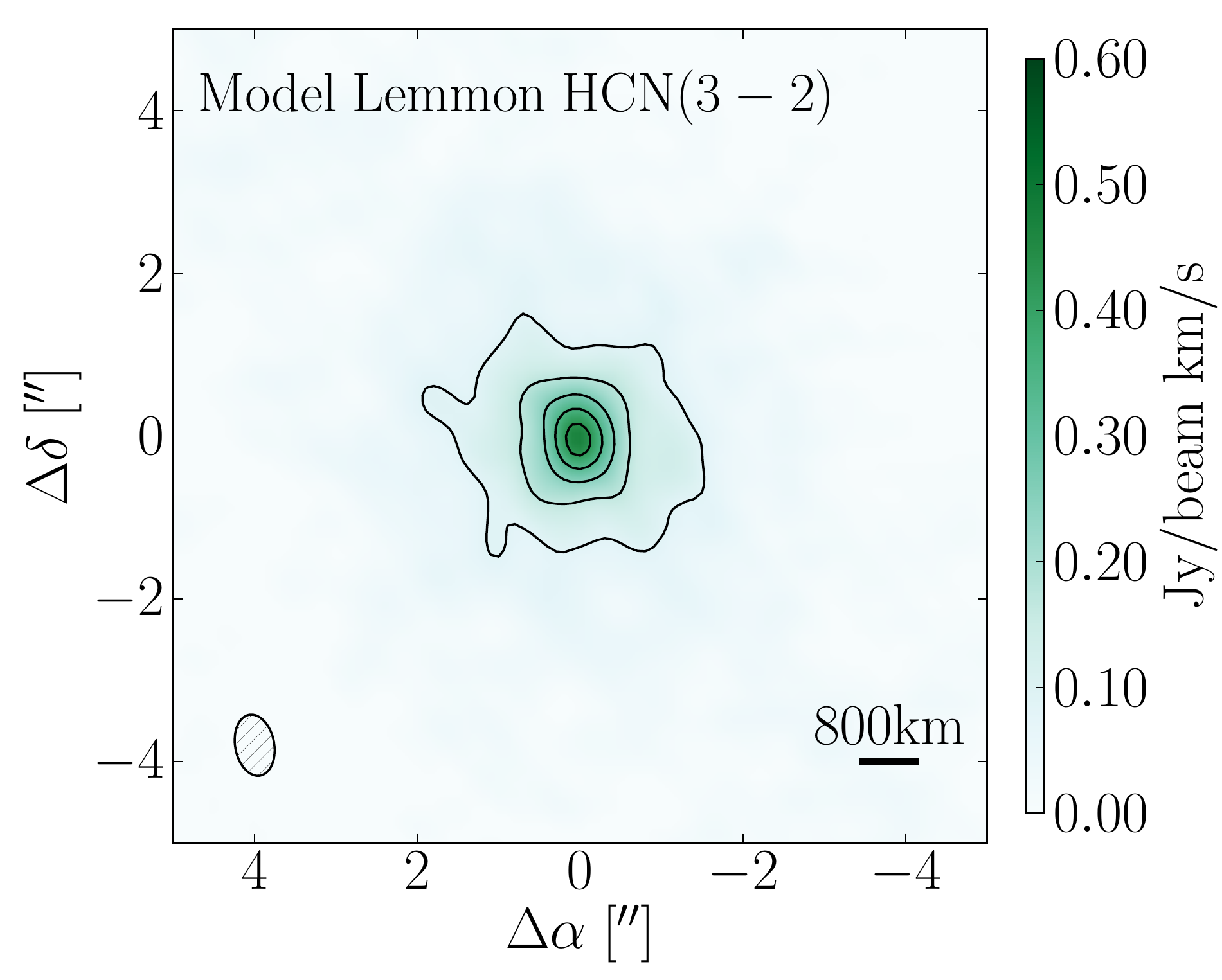}}  
	\subfigure{\label{} 
		\includegraphics[width=0.233\textwidth, trim={1.4cm 1.4cm 0 0.1cm}, clip]{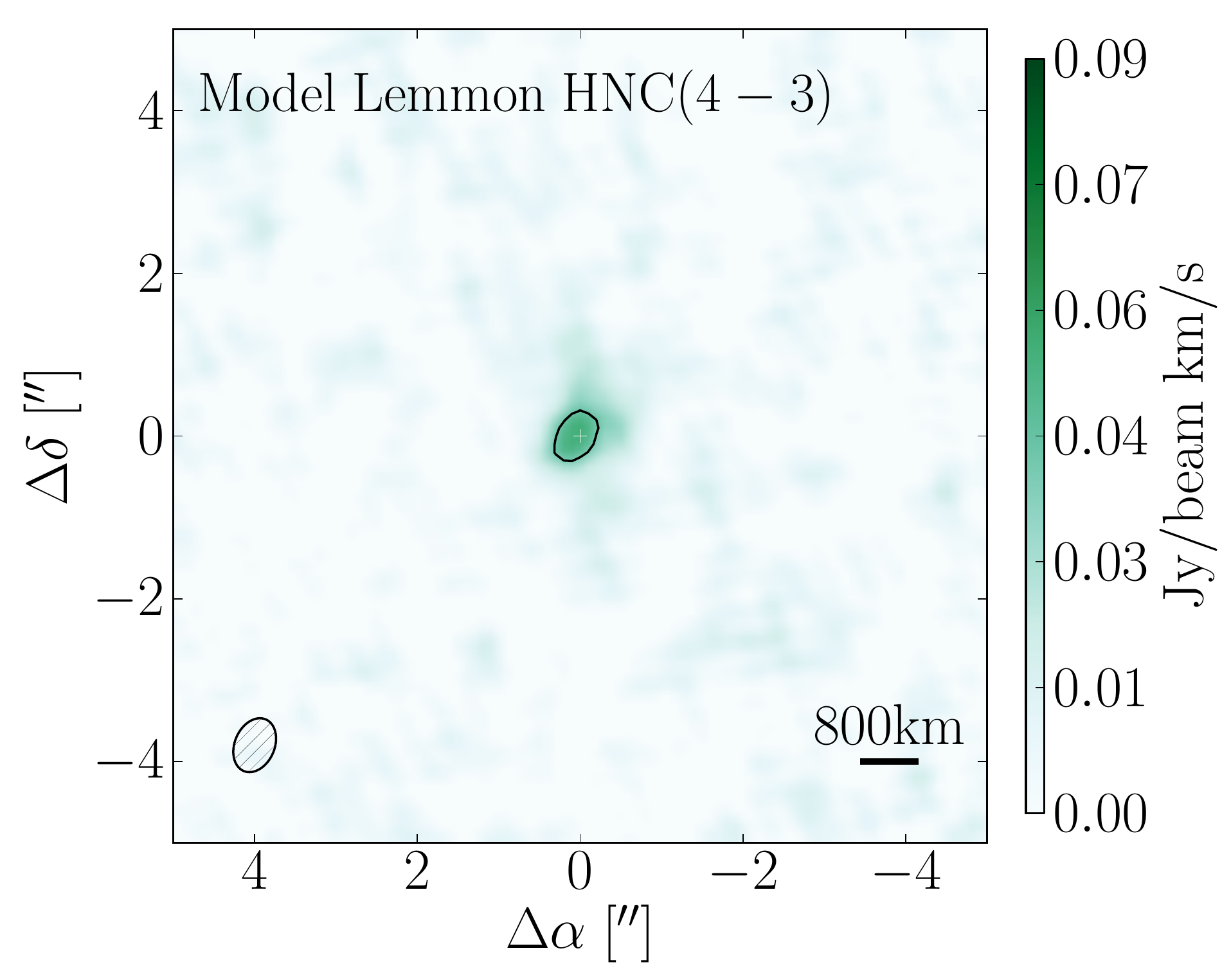}}
	\subfigure{\label{}
		\includegraphics[width=0.233\textwidth, trim={1.4cm 1.4cm 0 0.1cm}, clip]{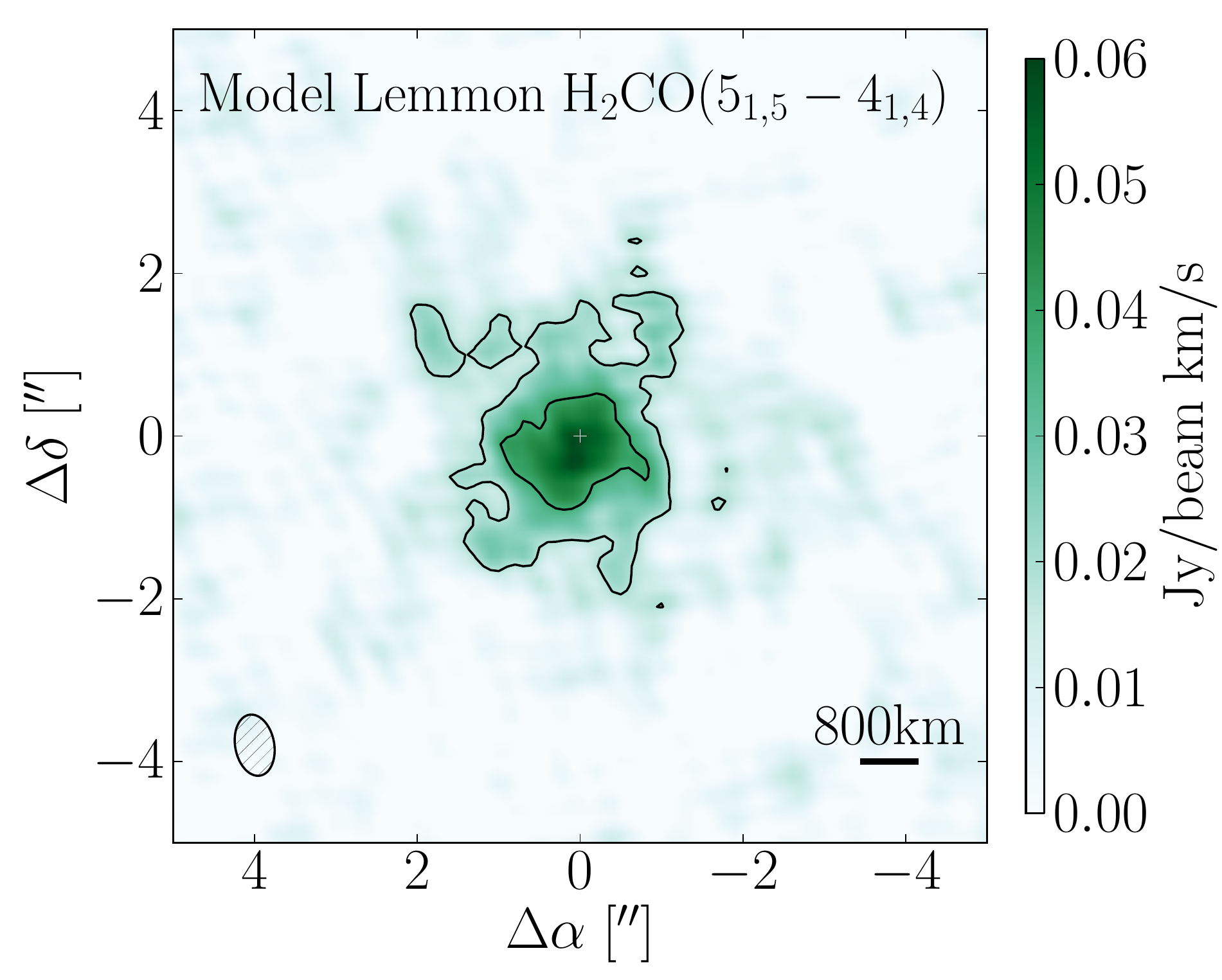}}
	
	\subfigure{\label{} 
		\includegraphics[width=0.248\textwidth, trim={0.3cm 1.4cm 0 0.4cm}, clip]{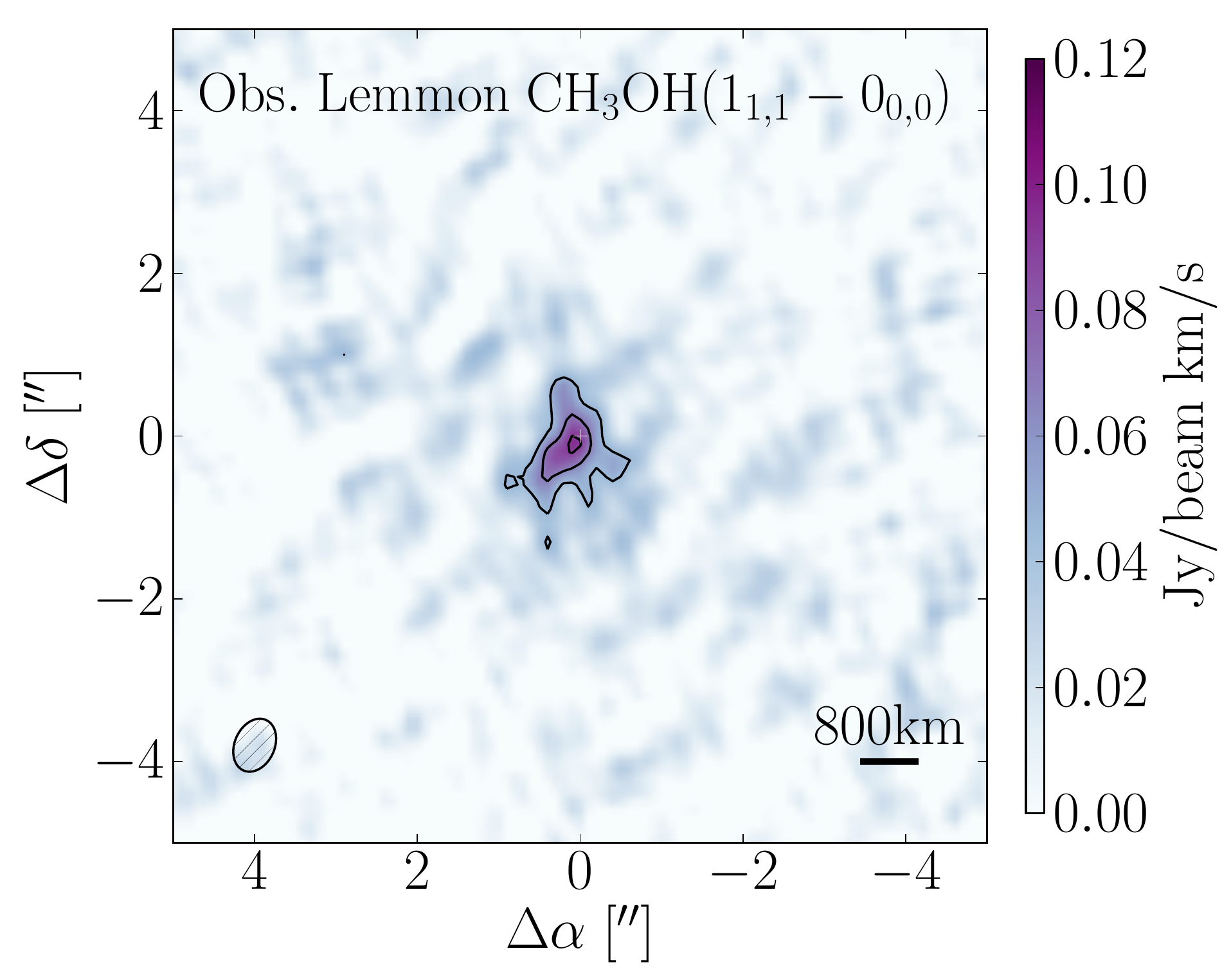}} 
	\subfigure{\label{}
		\includegraphics[width=0.233\textwidth, trim={1.4cm 1.4cm 0 0.1cm}, clip]{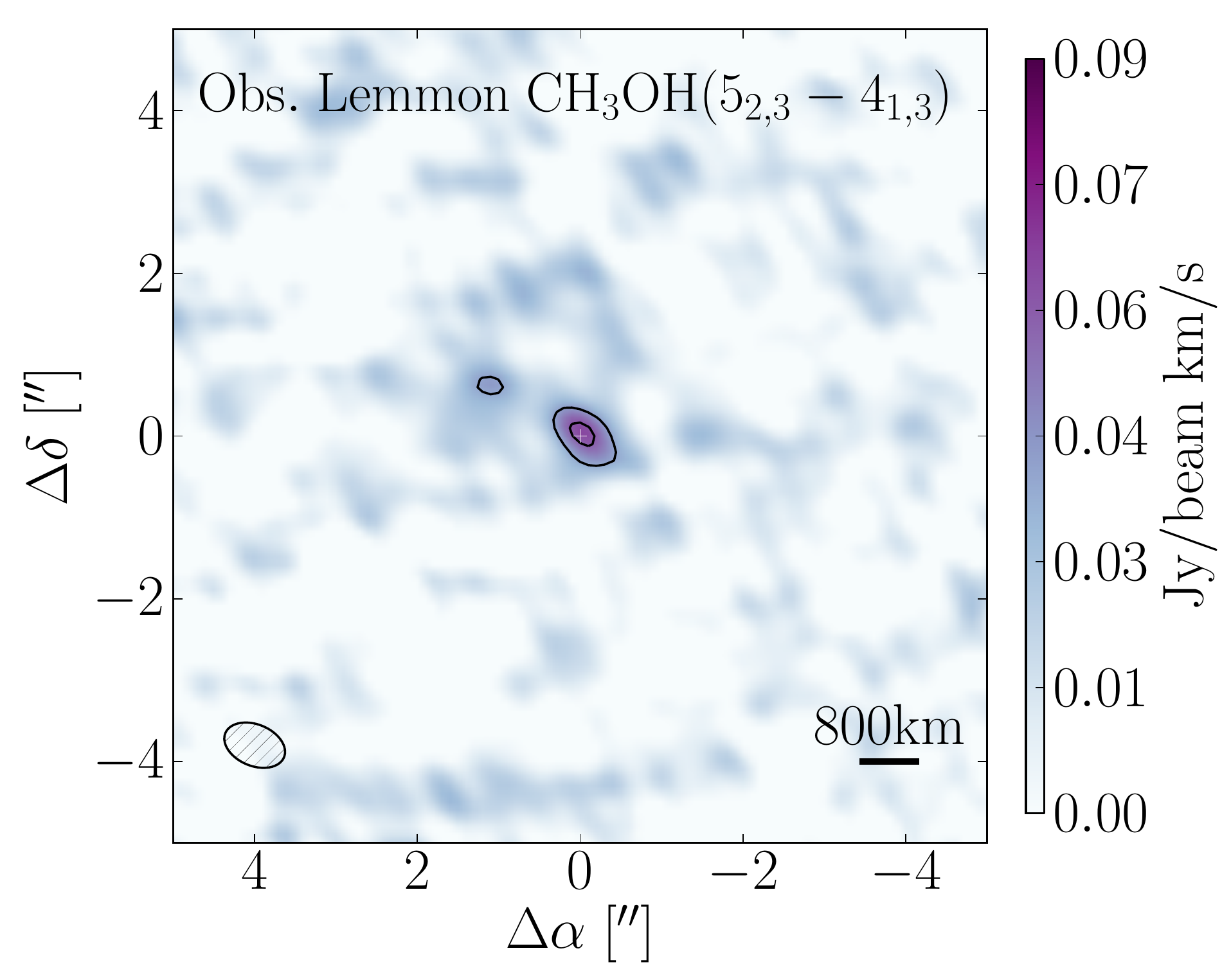}}  
	\subfigure{\label{} 
		\includegraphics[width=0.233\textwidth, trim={1.4cm 1.4cm 0 0.1cm}, clip]{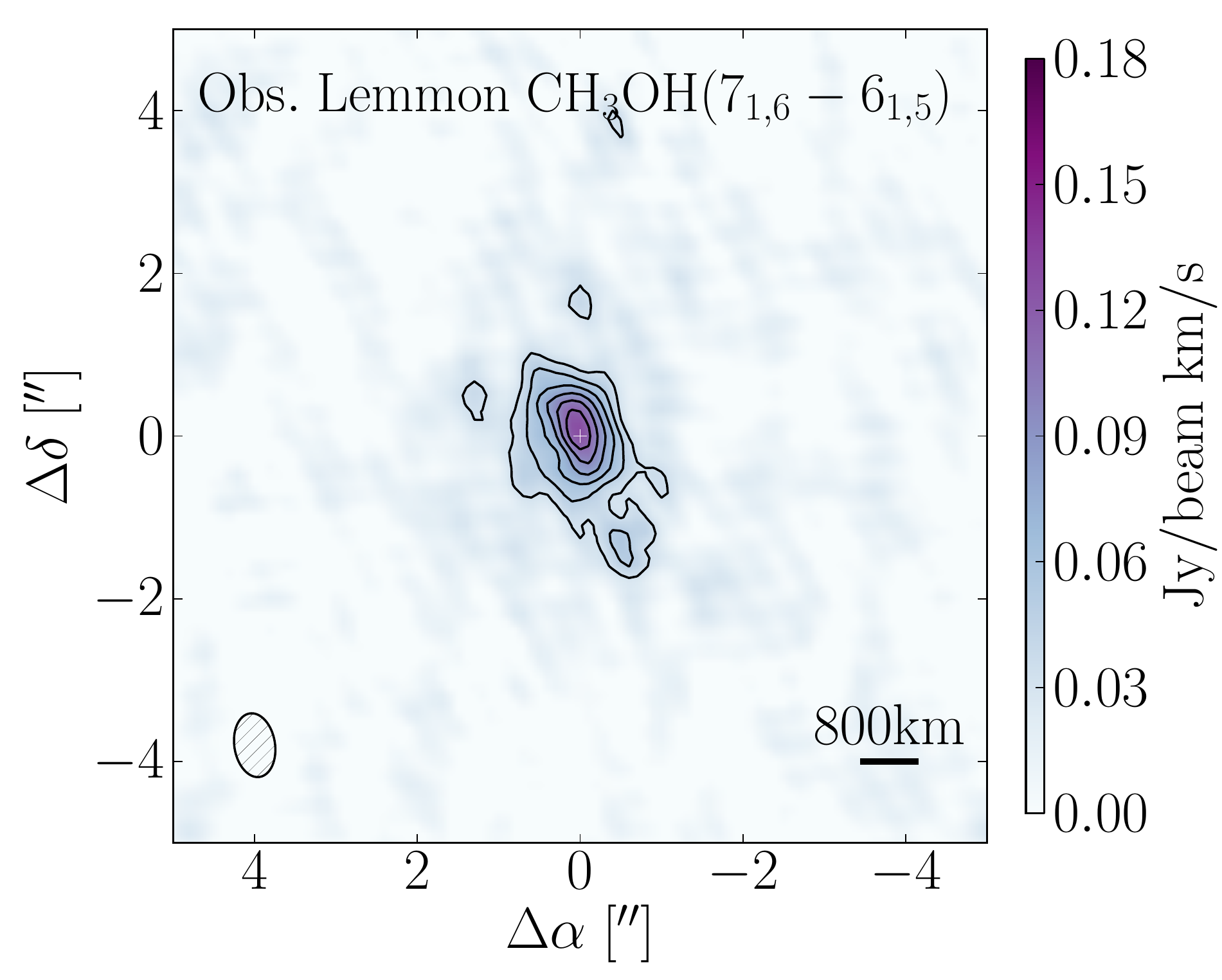}}
	\subfigure{\label{}
		\includegraphics[width=0.233\textwidth, trim={1.4cm 1.4cm 0 0.1cm}, clip]{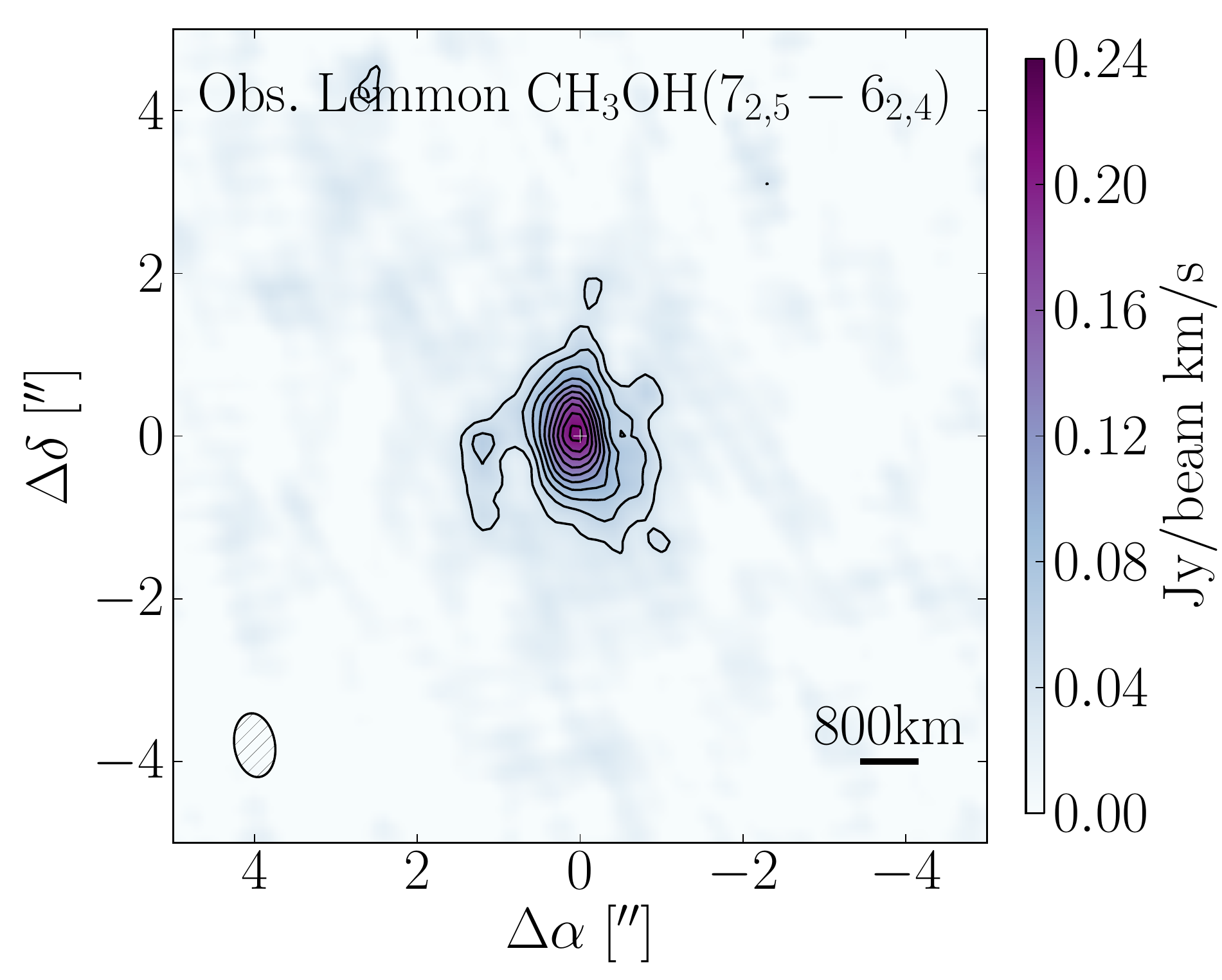}}	
	
	\subfigure{\label{}
		\includegraphics[width=0.248\textwidth, trim={0.3cm 0.3cm 0 0.4cm}, clip]{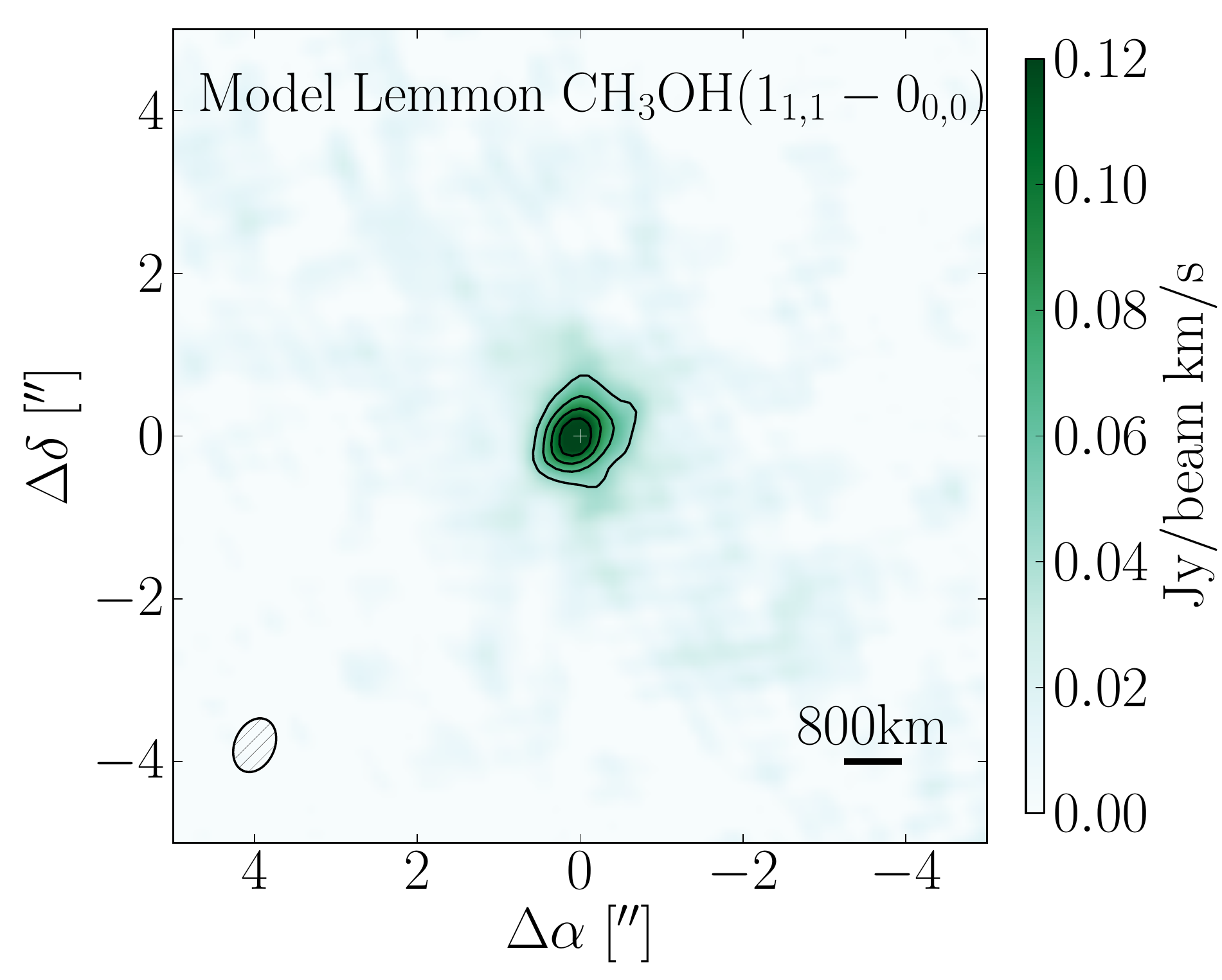}}	
	\subfigure{\label{}
		\includegraphics[width=0.233\textwidth, trim={1.3cm 0.3cm 0 0.2cm}, clip]{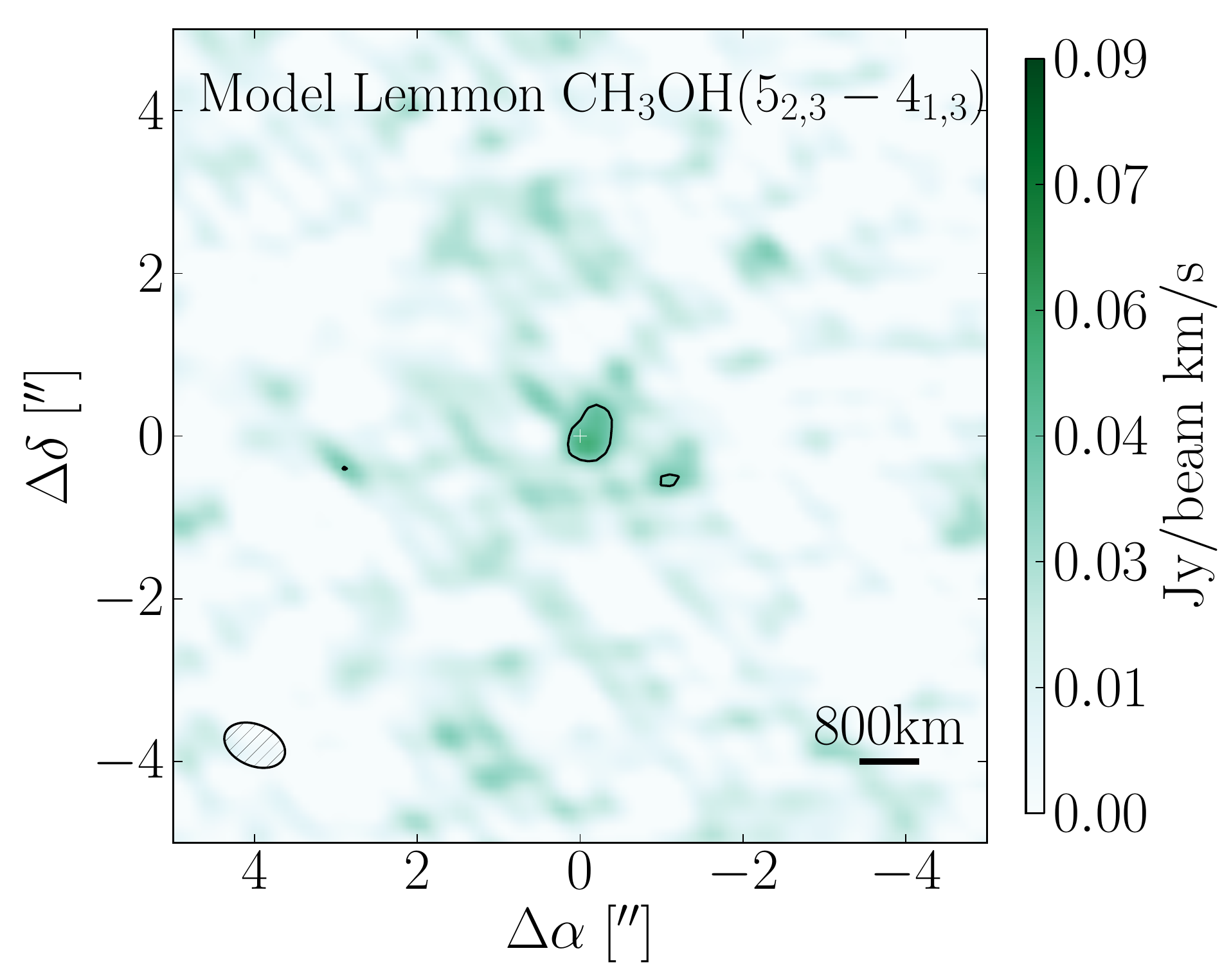}}
	\subfigure{\label{}
		\includegraphics[width=0.233\textwidth, trim={1.3cm 0.3cm 0 0.2cm}, clip]{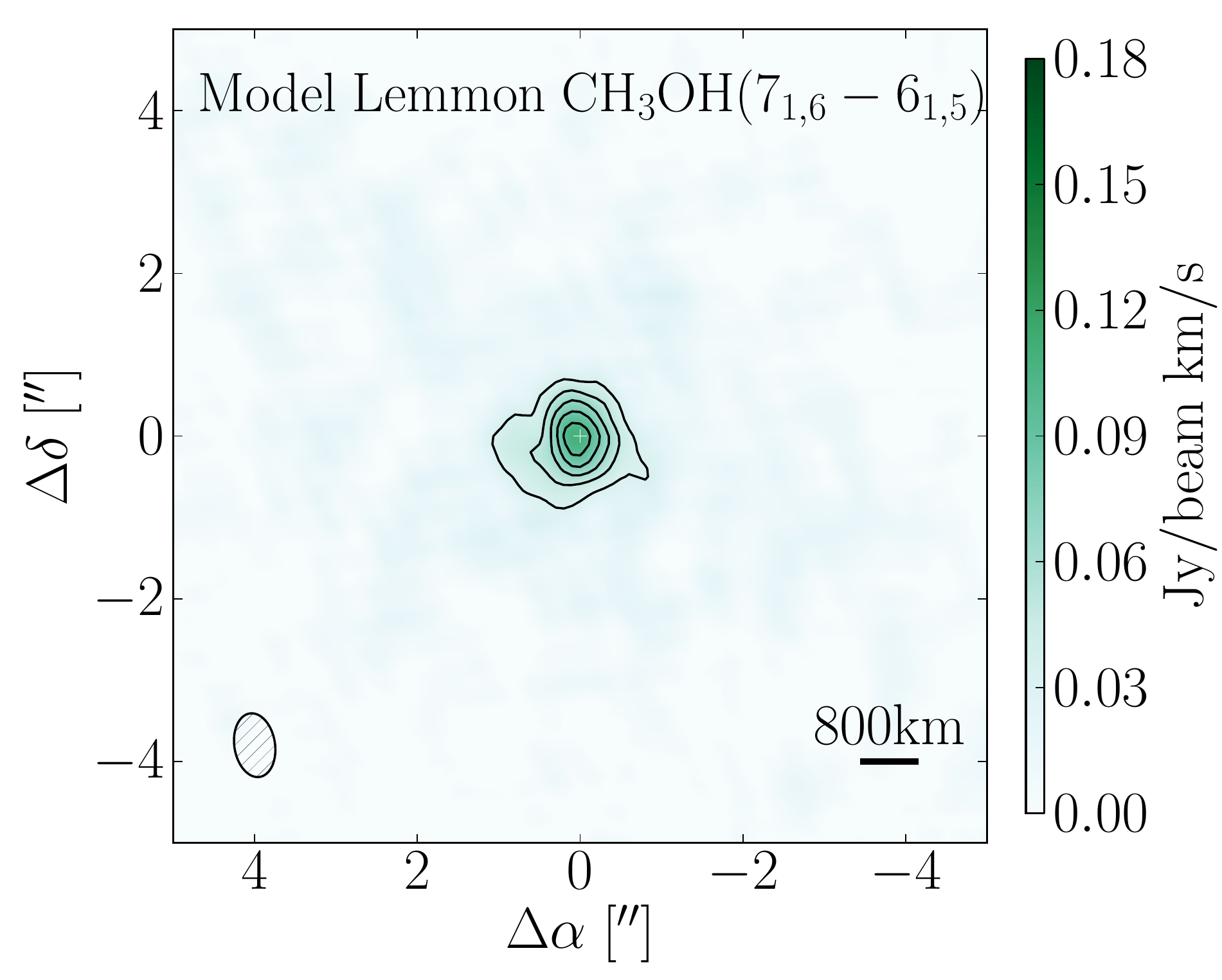}}
	\subfigure{\label{}
		\includegraphics[width=0.233\textwidth, trim={1.3cm 0.3cm 0 0.2cm}, clip]{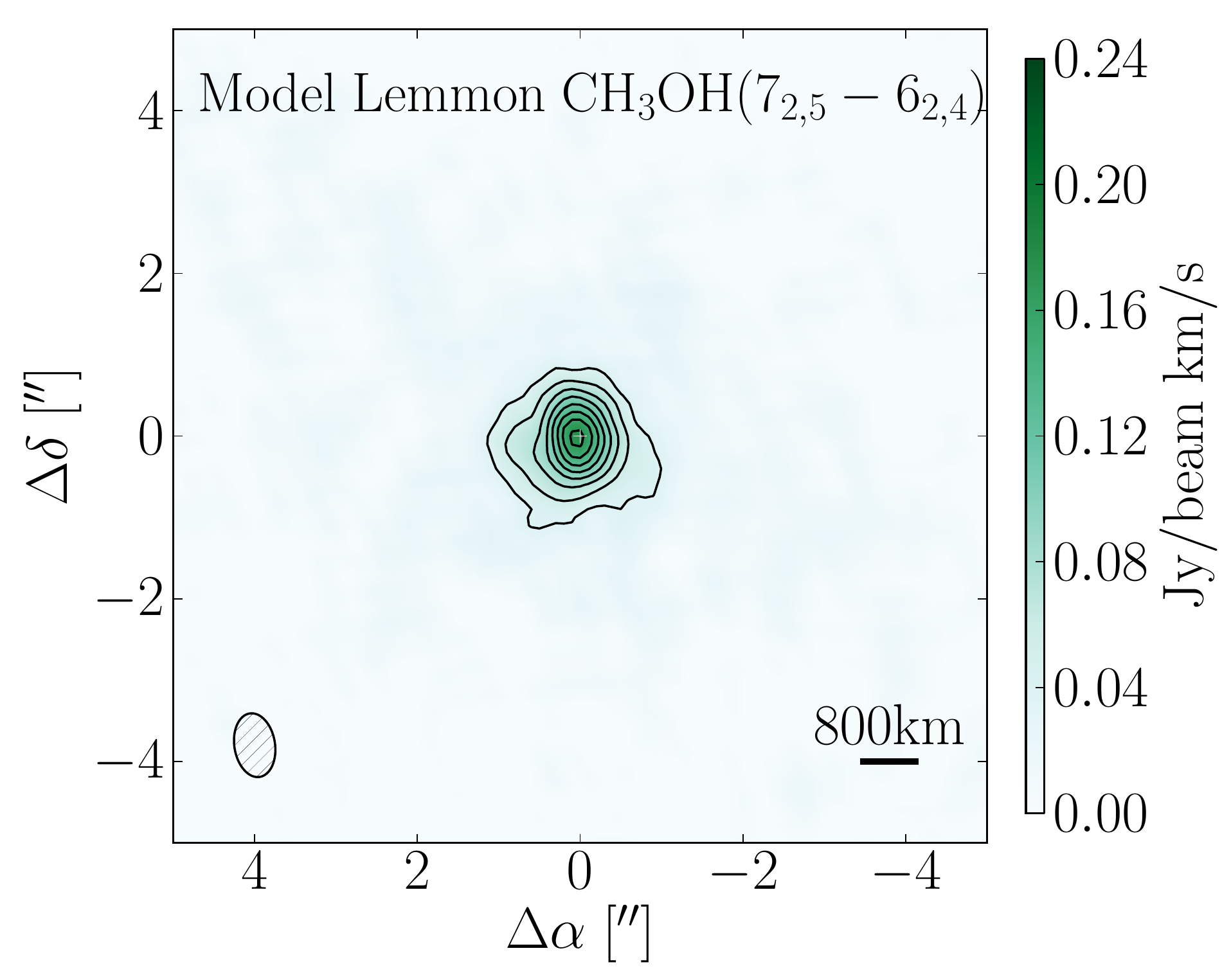}}

\caption[Lemmon Detections of HCN, HNC, H2CO and CH3OH]{Velocity integrated intensity maps (contours and colour) of HCN(4-3), HCN(3-2), HNC(4-3), H$_2$CO(5$_{1,5}$-4$_{1,4}$), CH$_3$OH(1$_{1,1}$-0$_{0,0}$), CH$_3$OH(5$_{2,3}$-4$_{1,3}$), CH$_3$OH(7$_{1,6}$-6$_{1,5}$) and CH$_3$OH(7$_{2,6}$-6$_{2,4}$) detected in the coma of comet Lemmon in blue and model counterparts (Section \ref{sec:model}) in green. Colours indicate intensity and contours are in steps of 30$\sigma$ for HCN(4-3), 10$\sigma$ for HCN(3-2), 3$\sigma$ for H$_2$CO(5$_{1,5}$-4$_{1,4}$), CH$_3$OH(1$_{1,1}$-0$_{0,0}$), CH$_3$OH(7$_{1,6}$-6$_{1,5}$) and CH$_3$OH(7$_{2,6}$-6$_{2,4}$), starting at 6$\sigma$ for CH$_3$OH(1$_{1,1}$-0$_{0,0}$), CH$_3$OH(7$_{1,6}$-6$_{1,5}$) and CH$_3$OH(7$_{2,6}$-6$_{2,4}$) and 1$\sigma$ for HNC(4-3) and CH$_3$OH(5$_{2,3}$-4$_{1,3}$) starting at 3$\sigma$ and 2$\sigma$ respectively where $\sigma$ is the RMS noise in each map. Crosses mark the peak continuum emission.}
\label{fig:Lemmon_B7}

\end{figure*}

\begin{figure*}
	\centering
\subfigure{\label{} 
	\includegraphics[width=0.248\textwidth, trim={0.3cm 1.4cm 0 0.4cm}, clip]{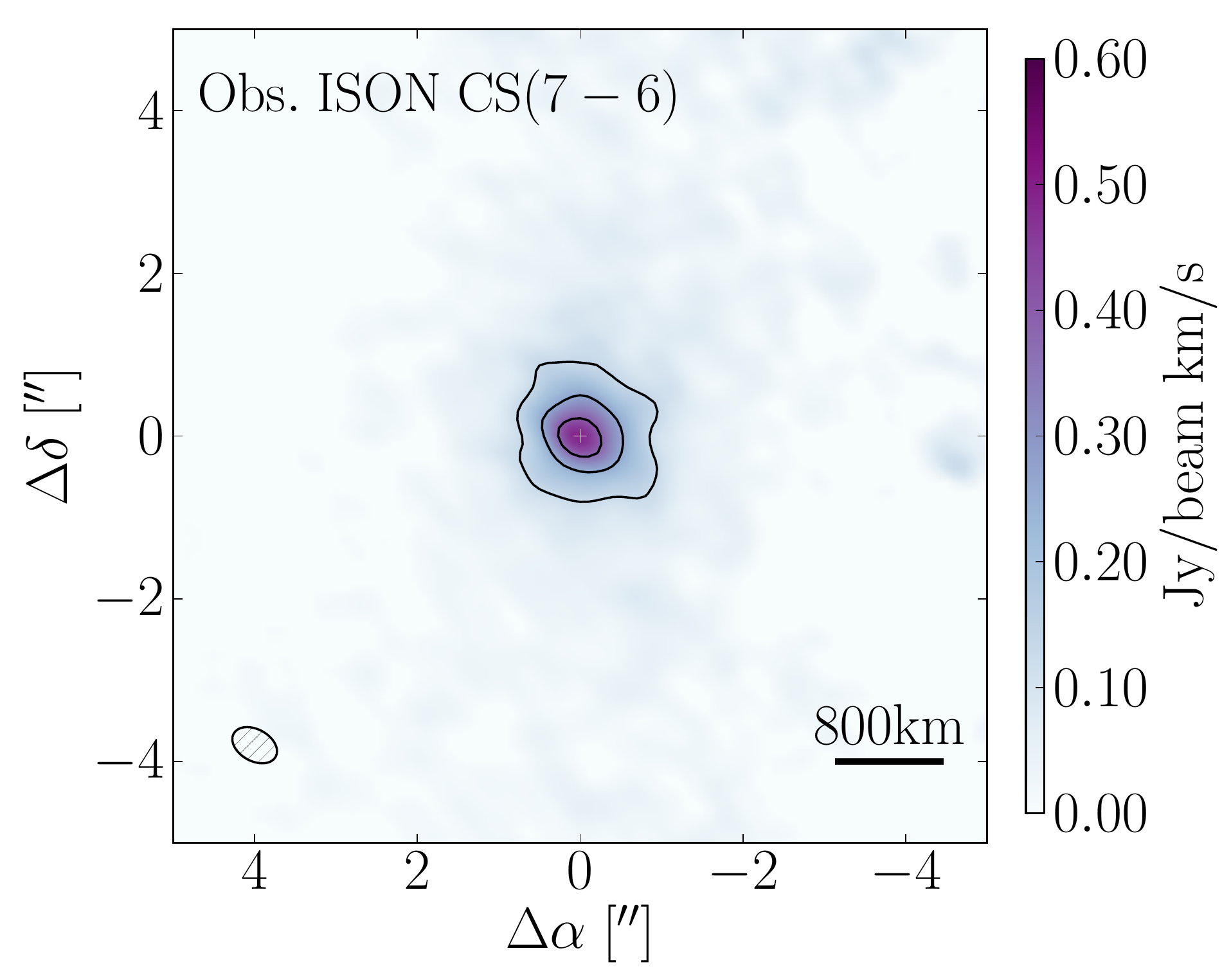}} 
\subfigure{\label{}
	\includegraphics[width=0.233\textwidth, trim={1.4cm 1.4cm 0 0.1cm}, clip]{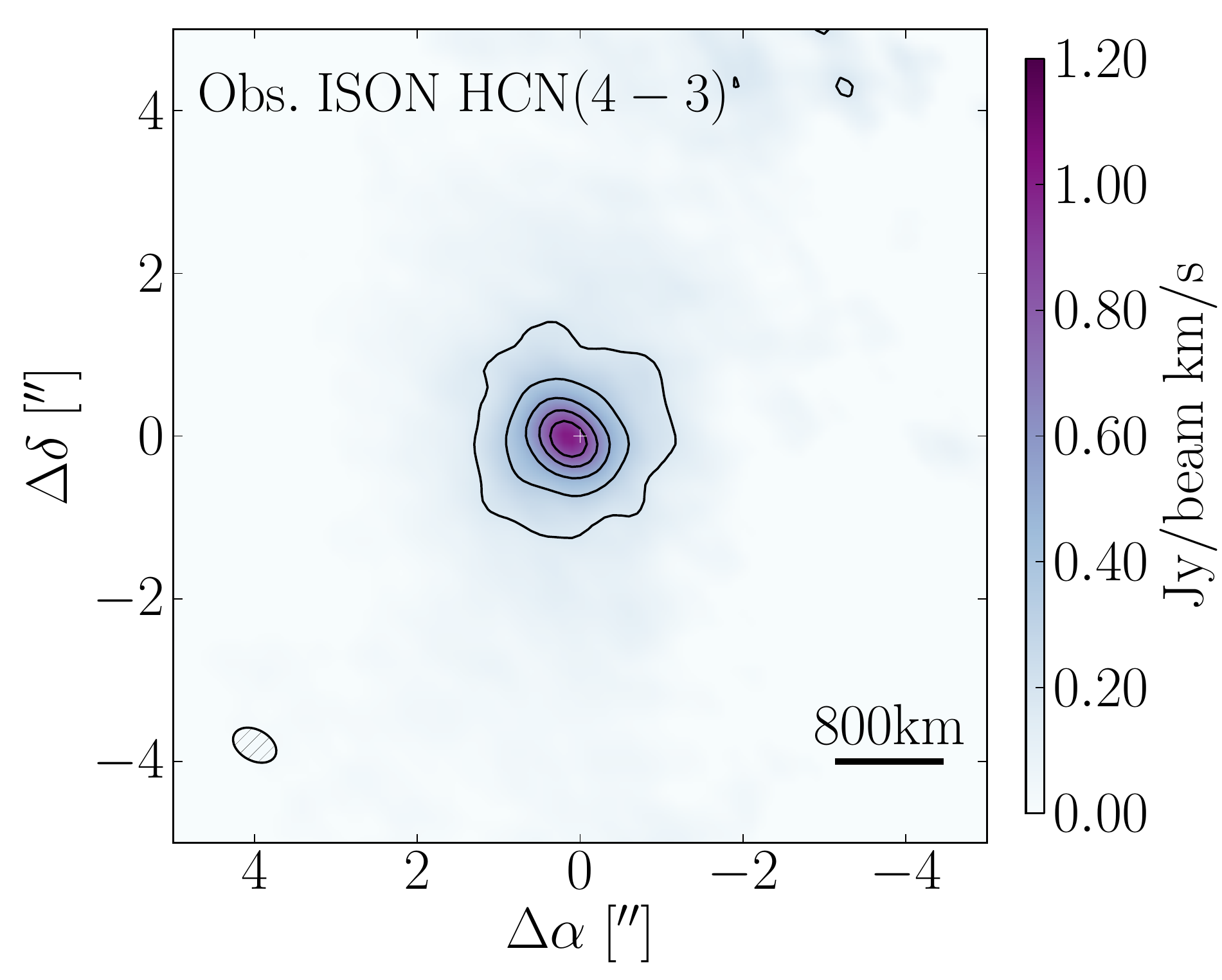}}  
\subfigure{\label{} 
	\includegraphics[width=0.233\textwidth, trim={1.4cm 1.4cm 0 0.1cm}, clip]{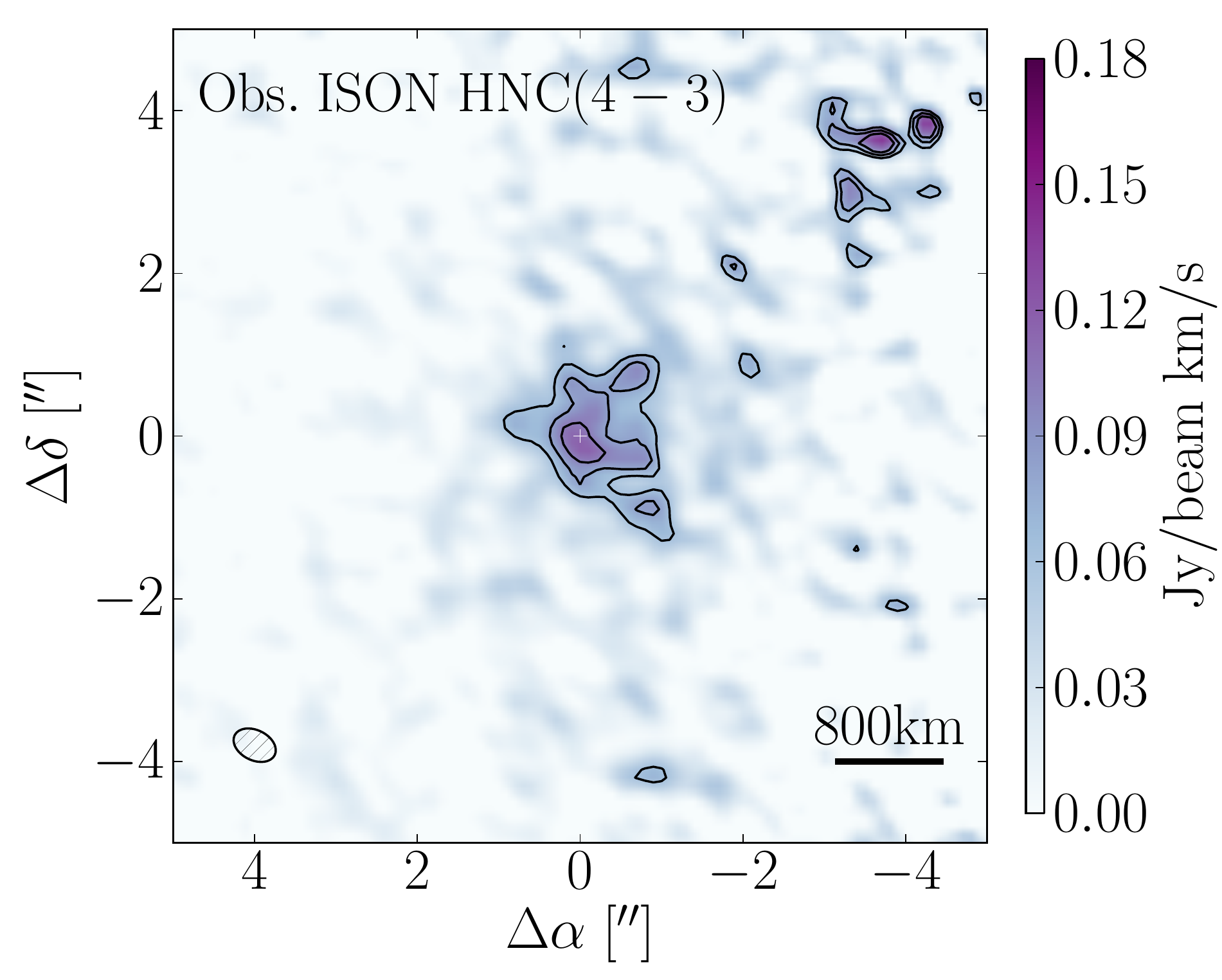}}
\subfigure{\label{}
	\includegraphics[width=0.233\textwidth, trim={1.4cm 1.4cm 0 0.1cm}, clip]{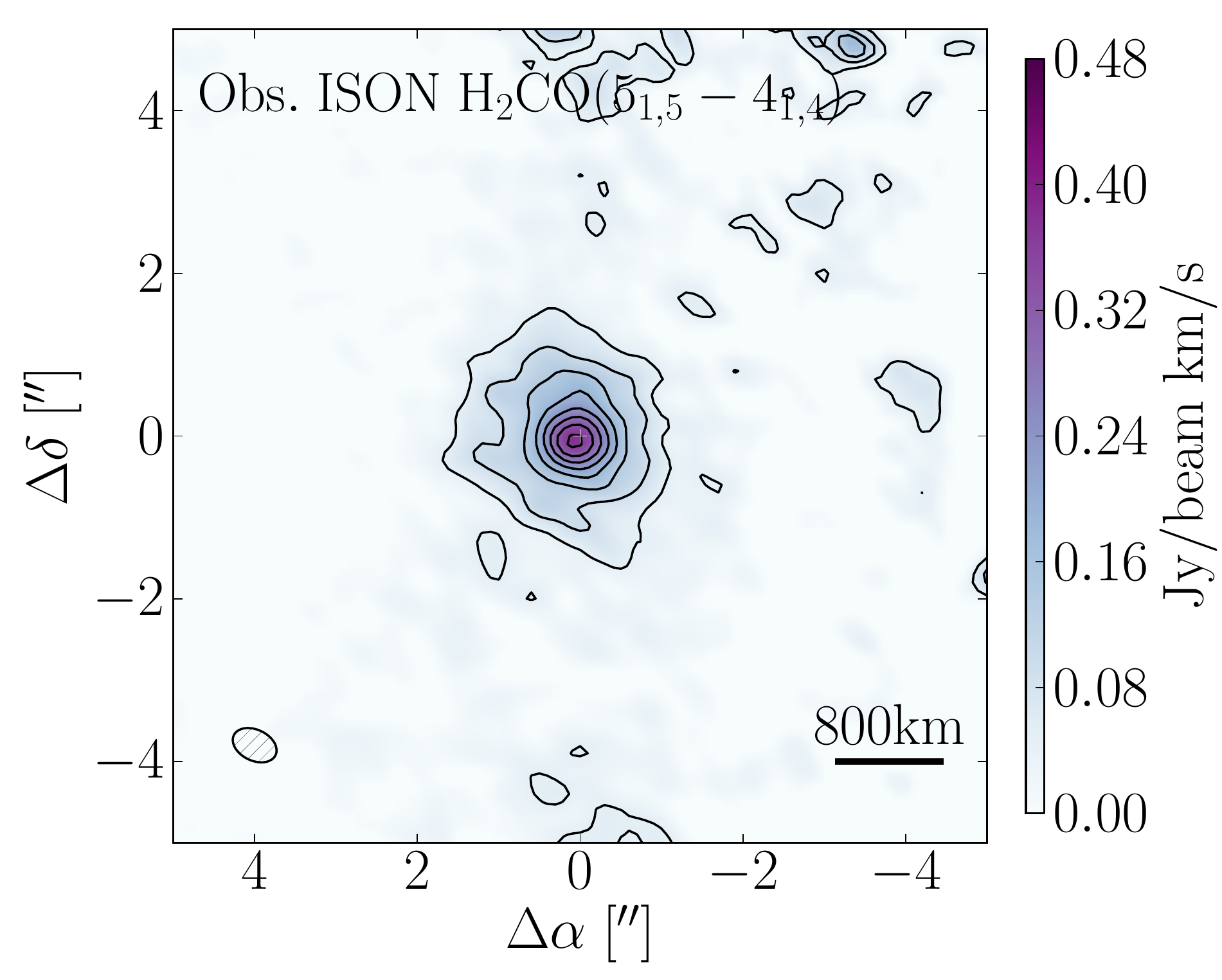}}
	
\subfigure{\label{} 
	\includegraphics[width=0.248\textwidth, trim={0.3cm 0.3cm 0 0.4cm}, clip]{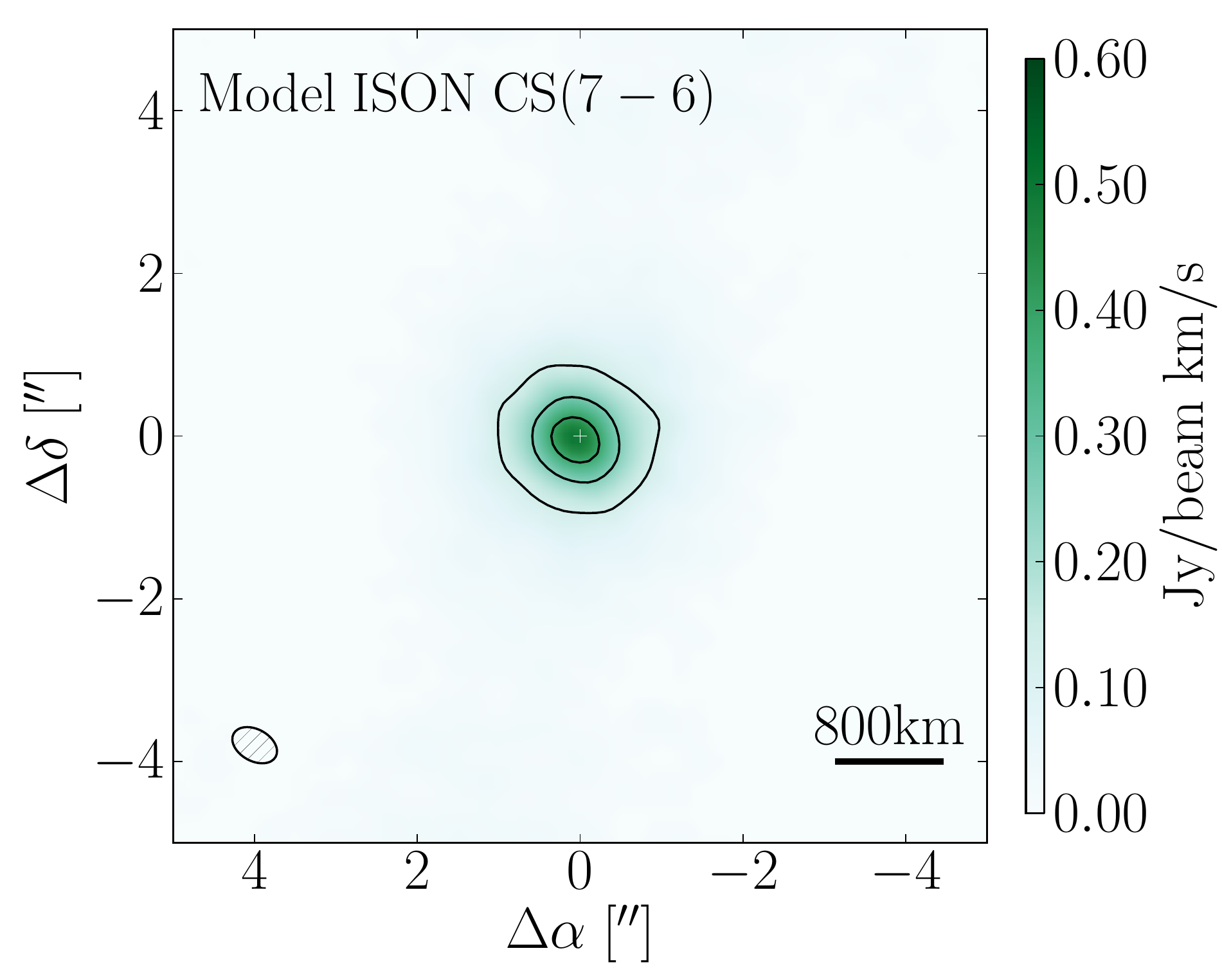}}
\subfigure{\label{}
	\includegraphics[width=0.233\textwidth, trim={1.4cm 0.3cm 0 0.1cm}, clip]{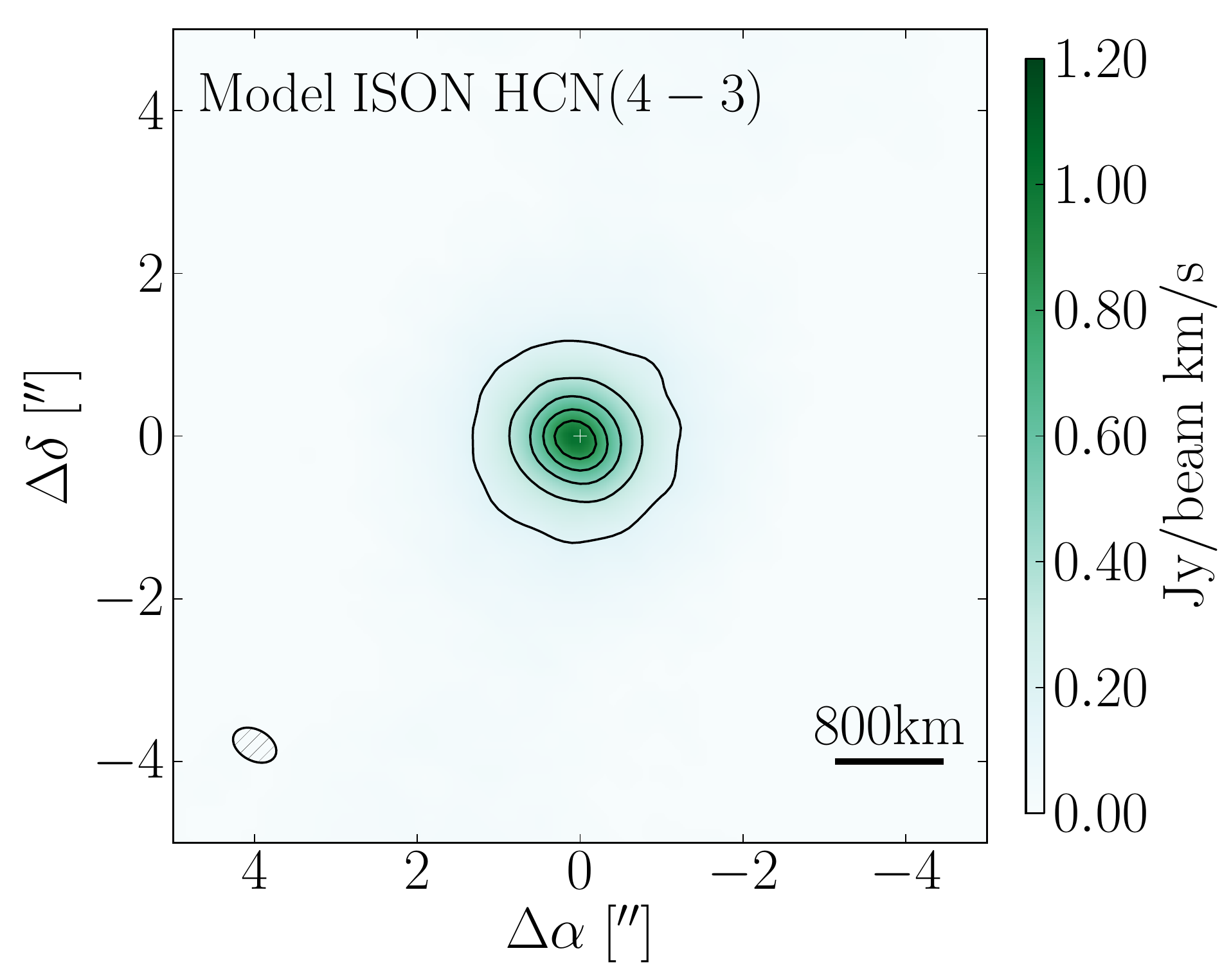}}  
\subfigure{\label{} 
	\includegraphics[width=0.233\textwidth, trim={1.4cm 0.3cm 0 0.1cm}, clip]{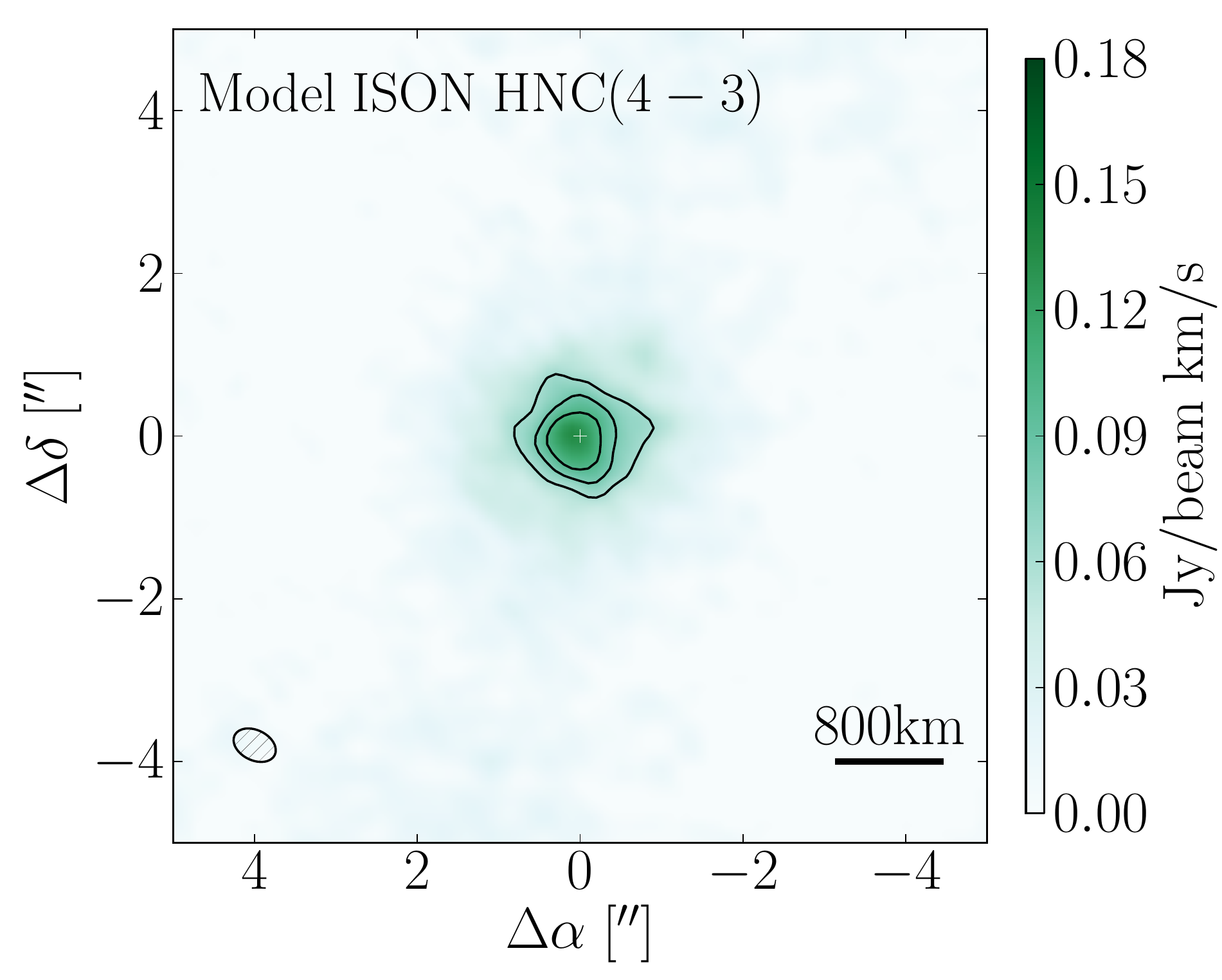}}
\subfigure{\label{}
	\includegraphics[width=0.233\textwidth, trim={1.4cm 0.3cm 0 0.1cm}, clip]{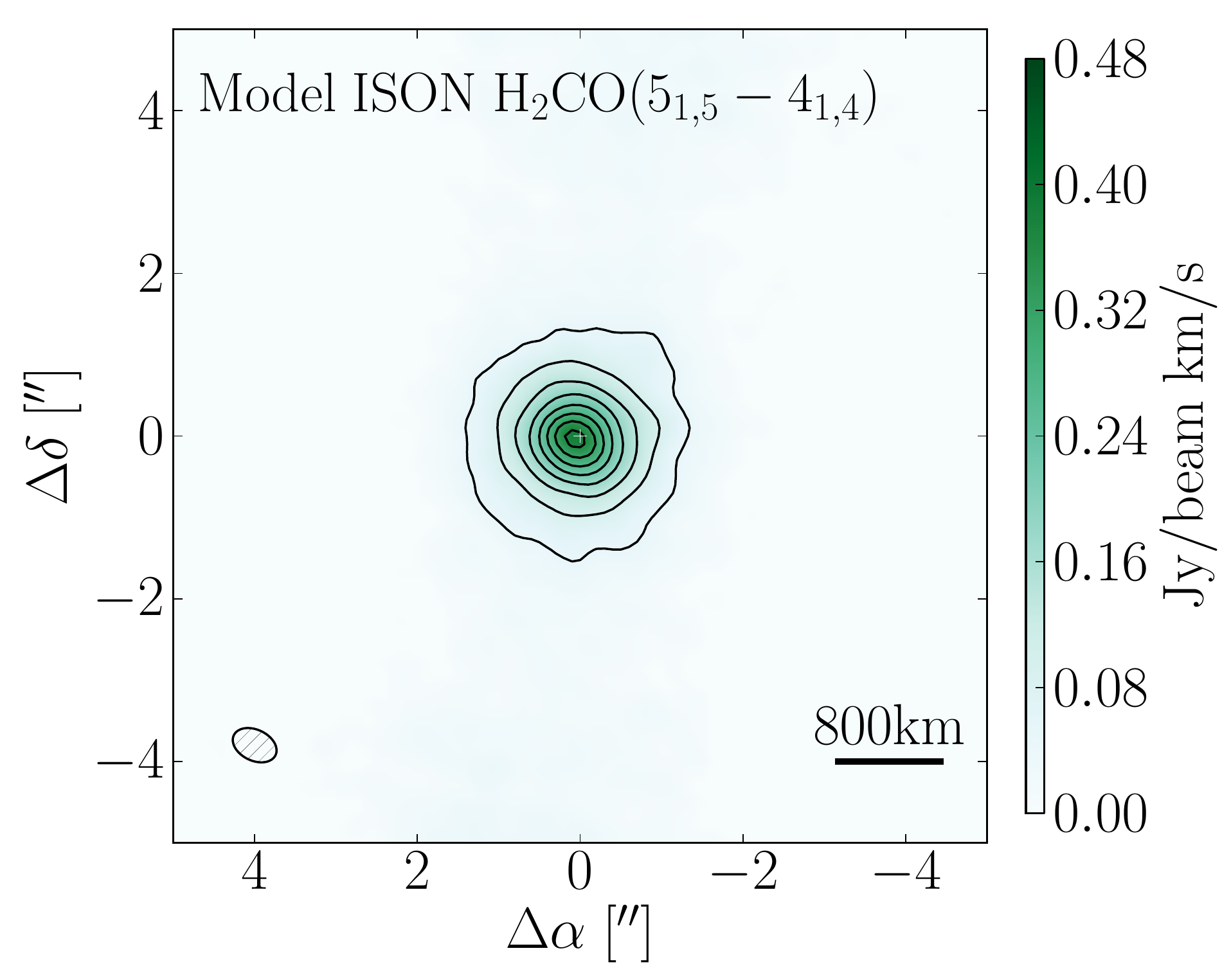}}	
	
\subfigure{\label{}
	\includegraphics[width=0.248\textwidth, trim={0.3cm 0.3cm 0 0.2cm}, clip]{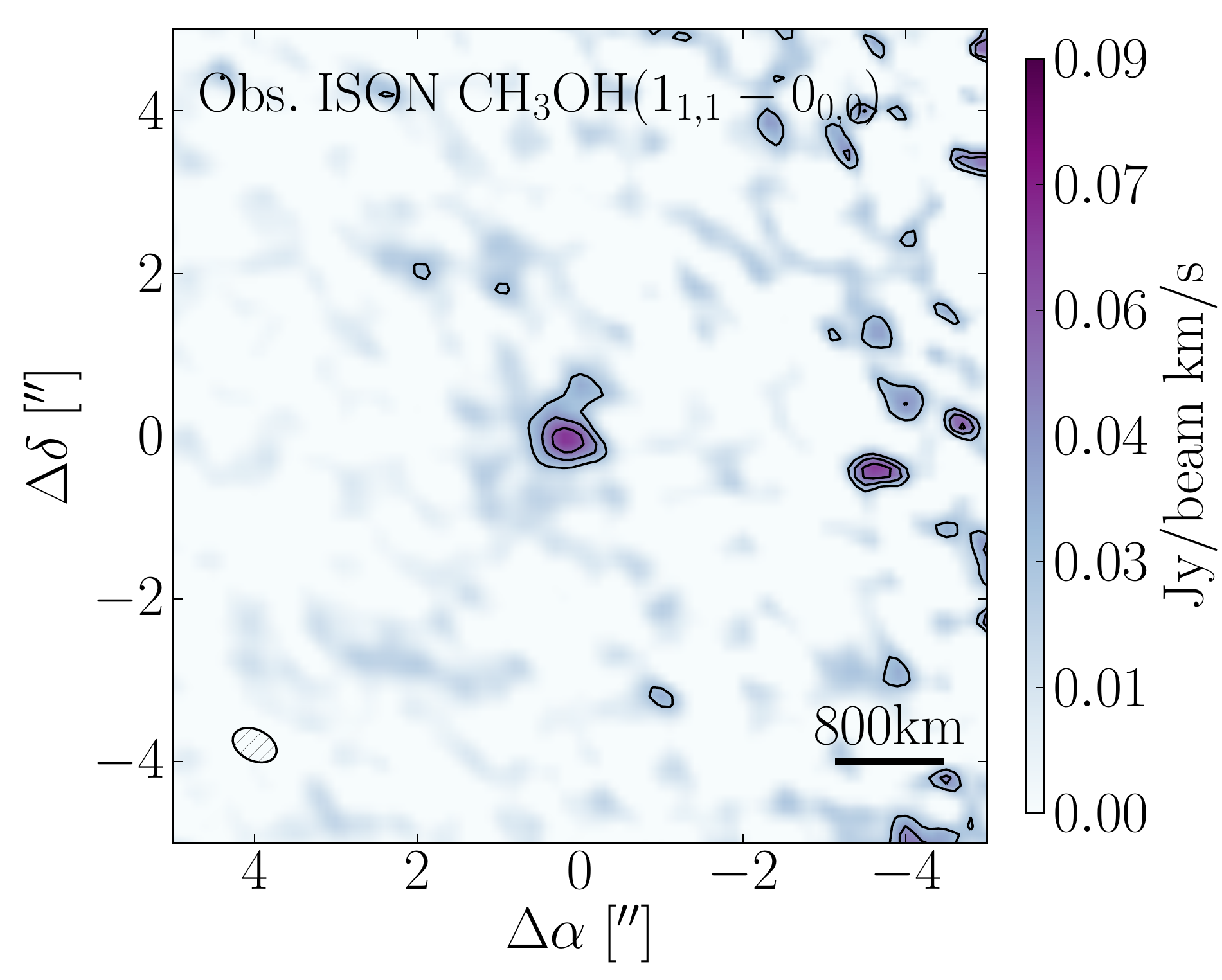}}	
\subfigure{\label{}
	\includegraphics[width=0.233\textwidth, trim={1.3cm 0.3cm 0 0.2cm}, clip]{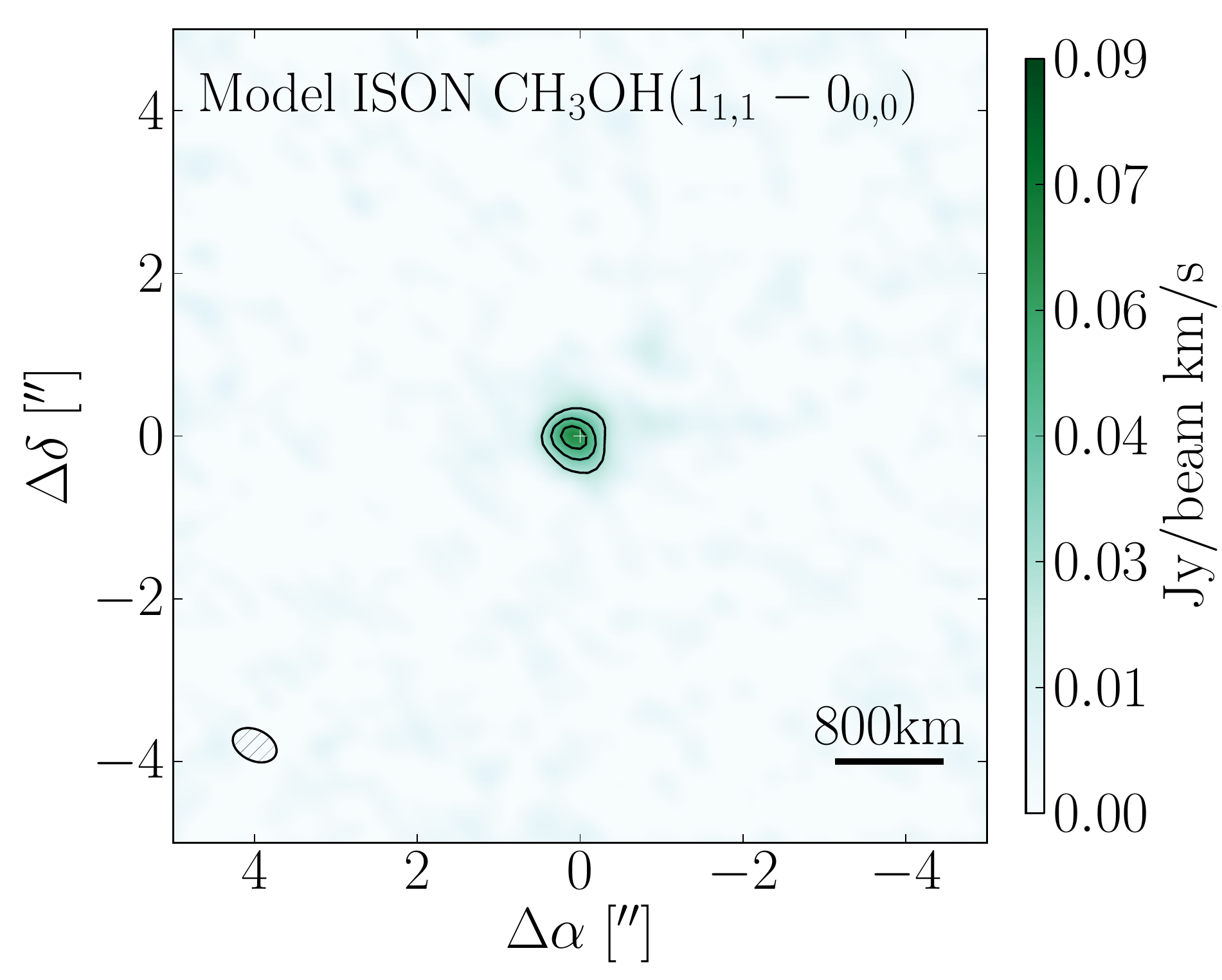}}	
	
\caption[ISON Band 7 detected species]{Velocity integrated intensity maps (contours and colour) of CS(7-6), HCN(4-3), HNC(4-3), H$_2$CO(5$_{1,5}$-4$_{1,4}$) and CH$_3$OH(1$_{1,1}$-0$_{0,0}$) detected in the coma of comet ISON in blue and model counterparts (Section \ref{sec:model}) in green. Colours indicate intensity and contours are in steps of 10$\sigma$ for CS(7-6) and HCN(4-3), 3$\sigma$ for H$_2$CO(5$_{1,5}$-4$_{1,4}$) and 1$\sigma$ for HNC(4-3) and CH$_3$OH(1$_{1,1}$-0$_{0,0}$) starting at 3$\sigma$ and 2$\sigma$ respectively where $\sigma$ is the RMS noise in each map. Crosses mark the peak continuum emission.}
\label{fig:ISON_B7}
\end{figure*}

We present the first AMLA detections of carbon monosulfide, CS, in comet ISON, as well as several methanol, CH$_3$OH, transitions and the line rations of HCN($J$=4--3)/($J$=3--2) in comet Lemmon. For completeness, and in order to compare our method to that reported by \cite{Cordiner2014}, we also present the detection of hydrogen cyanide, HCN, its metastable isomer hydrogen isocyanide, HNC, and formaldehyde, H$_2$CO. In contrast to \cite{Cordiner2017b}, who discuss the time variability of the HNC, H$_2$CO and CH$_3$OH emission in comet ISON, our imaging is time averaged and our derived production rates are time and spherically averaged by assuming a spatially uniform production rate. Spectrally integrated flux maps of all species, as well as their model counterparts, to be discussed in Section \ref{sec:model}, can be seen in Figs. \ref{fig:Lemmon_B7} and \ref{fig:ISON_B7}, while spectrally integrated peak fluxes, spectrally and spatially integrated total fluxes and molecular production rates are listed in Table \ref{tab:flux}.

Toward each comet a transition of HCO$^+$ was targeted in the observations but not detected in either. For completeness we therefore report 3$\sigma$ upper limits of 2.25$\times$10$^{-2}$ Jy beam$^{-1}$ km s$^{-1}$ for the velocity integrated peak intensity of HCO$^+$($J$=3-2), observed towards comet Lemmon, and 3.06 $\times$10$^{-2}$ Jy beam$^{-1}$ km s$^{-1}$ for HCO$^+$($J$=4-3), detected towards comet ISON. We do not model this emission because HCO$^+$ has an extended origin, probably a product of ion-neutral chemistry in the coma as seen in comets 67P/Churyumov-Gerasimenko \citep{Fuselier2015} and Hale-Bopp (1995 O1) \citep{Wright1998}, and ALMA has limited sensitivity to such extended emission.

\subsection{Comet Lemmon} \label{subsec:Lemmon}

Four species are detected towards comet Lemmon. HCN and CH$_3$OH show symmetric spatial distributions both of which peak at approximately the same location as the dust continuum, which we take as the location of the cometary nucleus, indicated by a cross in the figures. This is in agreement with HCN and CH$_3$OH being primary species, i.e., species released directly from the nucleus. While having the same general distribution, HCN and CH$_3$OH have very different velocity integrated line intensities, with that of HCN an order of magnitude higher than those of CH$_3$OH. In contrast to HCN and CH$_3$OH, HNC and H$_2$CO have much more distributed origins with similar integrated flux intensities, slightly below those of the CH$_3$OH transitions. The distributed origins indicate that HNC and H$_2$CO are either the result of gas-phase chemistry in the cometary coma, or that they are transported away from the nucleus by some refractory compound before being evaporated. In Sections \ref{subsec:molecular_production_rates} and \ref{subsec:chemistry} we will discuss the parent scale lengths and production rates we derive for HNC, H$_2$CO and CH$_3$OH in detail and how these compare with the various formation routes discussed above.

\subsection{Comet ISON} \label{subsec:ISON} 

Towards comet ISON we detect five molecular species. As is the case for comet Lemmon, HCN and CH$_3$OH show centrally peaked distributions coinciding with the peak of the continuum. We see the same trend in integrated line flux in comet ISON as in comet Lemmon, with the HCN transition being brighter than that of CH$_3$OH by more than an order of magnitude. The spatial distributions of HNC and H$_2$CO are again more distributed compared to HCN and CH$_3$OH but to a lesser extent than in the case of comet Lemmon. This may be due to the generally higher activity, and consequently higher molecular production rates, of comet ISON compared to comet Lemmon. A higher production rate is consistent with the heliocentric distance of ISON being only one third of the heliocentric distance of comet Lemmon. A less extended spatial distribution would also imply that new molecules are released from the nucleus, or produced in the coma of comet ISON, faster than they are transported away. The integrated line intensities of HNC and H$_2$CO are lower than that of HCN but, in contrast to comet Lemmon, higher than for CH$_3$OH.

In addition to the species discussed above, we report the first detection of CS by ALMA in the coma of comet ISON. We find the integrated line flux of CS to be half that of HCN but greater than those of HNC, H$_2$CO and CH$_3$OH. CS was first detected in comet West (1975 VI) \citep{Smith1980} through ultraviolet spectroscopy but has since been observed in a number of other objects, at a number of different wavelengths \cite[see][for a summay]{Mumma2011}. In Section \ref{subsec:molecular_production_rates} we derive the production rate and parent length scale for CS and in Section \ref{subsec:chemistry} we will discuss the possible formation scenarios for this molecule.

\begin{table*} 
	\begin{small}	
		\centering
		\caption{Integrated Peak and Total Flux}
		\label{tab:flux}
		\begin{tabular}{ccccccccccc}
			\toprule
			Source & Species  & Transition & E$_{\mathrm{up}}$ & Q\tablefootmark{a} &Q$_{\mathrm{X}}$/Q$_{\mathrm{H_2O}}$ & L$_{\mathrm{p}}$ & \multicolumn{2}{c}{Peak Flux} & \multicolumn{2}{c}{Total Flux\tablefootmark{b}} \\
			&&& [K] & [10$^{26}$ s$^{-1}$] & [$\%$] & [km] & \multicolumn{2}{c}{[Jy beam$^{-1}$ km s$^{-1}$]} & \multicolumn{2}{c}{[Jy km s$^{-1}$]} \\
			\cmidrule(l){8-9}
			\cmidrule(l){10-11}
			&&&&&&& Obs.\tablefootmark{c} & Model & Obs.\tablefootmark{c} & Model \\
			\midrule    
			\multirow{8}{*}{\shortstack{C/2012 F6\\(Lemmon)}}
			& HCN			& 3-2 & 25.52 & 2.0 $\substack{+0.4 \\ -0.1}$ & 0.064 $\substack{+0.013 \\ -0.003}$ & 10 & 0.47$\pm$0.05 & 0.42 & 3.57$\pm$0.08 & 4.24 \\    
			&  		        & 4-3 & 42.53 & 2.3 $\substack{+0.7 \\ -0.1}$ & 0.130 $\substack{+0.040 \\ -0.006}$ & 10 & 1.13$\pm$0.11 & 1.03 & 8.63$\pm$0.12 & 8.71 \\
			& HNC           & 4-3 & 43.51 & 0.06 $\substack{+0.04 \\ -0.01}$ & 0.003 $\substack{+0.002 \\ -0.001}$ & 10 & 0.07$\pm$0.02 & 0.06 & 0.14$\pm$0.06\tablefootmark{d} & 0.22\tablefootmark{d} \\
			& HCO$^+$       & 3-2 & 25.68 & - & - & - & 2.25$\times$10$^{-2}$\tablefootmark{e} & - & - & - \\    
			& H$_2$CO       & 5$_{1,5}$-4$_{1,4}$ & 62.45 & 1.8 $\pm$0.3 & 0.102 $\pm$0.017 & 1400$\pm$300 & 0.06$\pm$0.01 & 0.06 & 0.79$\pm$0.04 & 0.78\\
			& CH$_3$OH      & 1$_{1,1}$-0$_{0,0}$ & 16.84 & 17.1 $\pm$0.9\tablefootmark{f} & 0.967 $\pm$0.053 & 10 & 0.10$\pm$0.01 & 0.14 & 0.80$\pm$0.06 & 0.88 \\
			&               & 5$_{2,3}$-4$_{1,3}$ & 57.07 & 12.0 $\substack{+11.0 \\ -2.0}$ & 0.384 $\substack{+0.352 \\ -0.064}$ & 10 & 0.07$\pm$0.02 & 0.06 & 0.42$\pm$0.13 & 0.55 \\
			& 		        & 7$_{1,6}$-6$_{1,5}$ & 80.09 & 17.1 $\pm$0.9\tablefootmark{f} & 0.967 $\pm$0.053 & 10 & 0.13$\pm$0.01 & 0.11 & 0.88$\pm$0.04 & 0.76 \\ 
			&               & 7$_{2,5}$-6$_{2,4}$ & 87.26 & 17.1 $\pm$0.9\tablefootmark{f} & 0.967 $\pm$0.053 & 10 & 0.21$\pm$0.02 & 0.18 & 1.28$\pm$0.05 & 1.07 \\ 
			\midrule
			\multirow{5}{*}{\shortstack{C/2012 S1\\(ISON)}}
			& CS            & 7-6 & 65.83 & 6.7 $\substack{+0.8 \\ -0.7}$ & 0.191 $\substack{+0.023 \\ -0.020}$ & 200 $\pm$50 & 0.50$\pm$0.05 & 0.50 & 5.21$\pm$0.12 & 5.22 \\
			& HCN           & 4-3 & 42.53 & 4.0 $\pm$0.5 & 0.114 $\pm$0.014 & 150 $\pm$50 & 1.01$\pm$0.10 & 1.02 & 11.78$\pm$0.18 & 11.79 \\ 
			& HNC           & 4-3 & 43.51 & 1.8 $\substack{+0.1 \\ -0.5}$ & 0.051 $\substack{+0.003 \\ -0.014}$ & 1200 $\substack{+500 \\ -100}$ & 0.12$\pm$0.02 & 0.14 & 2.33$\pm$0.19 & 2.22 \\
			& HCO$^+$       & 4-3 & 42.86 & - & -  & - & 3.06$\times$10$^{-2}$\tablefootmark{e} & - & - & - \\    
			& H$_2$CO       & 5$_{1,5}$-4$_{1,4}$ & 62.45 & 8.0 $\pm$1.0 & 0.229 $\pm$0.029 & 250 $\pm$50 & 0.38$\pm$0.04 & 0.39 & 3.70$\pm$0.14 & 3.72\\
			& CH$_3$OH     & 1$_{1,1}$-0$_{0,0}$ & 16.84 & 17.0 $\pm$5.0 & 0.486 $\pm$0.143 & 10 & 0.07$\pm$0.02 & 0.07 & 0.25$\pm$0.13 & 0.24 \\    
			\bottomrule
		\end{tabular}
		\tablefoot{\tablefoottext{a}{Production rate including 1$\sigma$ error}. \tablefoottext{b}{Spectrally integrated line flux in circular aperture of 5 arcsec diameter centered on comet}. \tablefoottext{c}{Errors assume a 10$\%$ absolute flux calibration error}. \tablefoottext{d}{Spectrally integrated line flux in circular aperture of 3 arcsec diameter centered on comet}. \tablefoottext{e}{3$\sigma$ upper limit}. \tablefoottext{f}{Weighted average}.}
	\end{small}
\end{table*}

\section{Model} \label{sec:model}

To calculate molecular production rates we model the emission of each of the detected species. This is done using LIME \citep{Brinch2010}, a code for non-LTE line excitation and radiative transfer. We assume a spherically symmetric model with constant outflow velocity \citep{Haser1957,Combi2004} to describe the number density of molecules released from the cometary nucleus, i.e. parent molecules, ($n_{ \rm p}$), as a function of distance from the cometary nuclei ($r$) 

 \begin{equation} \label{eq:haser_parent_model}
 n_{ \rm p}(r) = \frac{Q}{4 \pi v_{\mathrm{exp}}r^2} \mathrm{exp} \left( -\frac{r \beta}{v_{\mathrm{exp}}} \right),
 \end{equation}
 
\noindent with $Q$ denoting the molecular production rate, $v_{\rm{exp}}$ the expansion velocity and $\beta$ the molecular photodestruction rate. 

Species produced by the destruction of parent molecules, referred to as daughter molecules, are described by

 \begin{equation} \label{eq:haser_daughter_model}
 n_{ \rm d}(r) = \frac{Q}{4 \pi v_{\mathrm{exp}}r^2} \frac{\frac{v_{\mathrm{exp}}}{\beta_{ \rm d}}}{\frac{v_{\mathrm{exp}}}{\beta_{ \rm d}}-L_{\mathrm{p}}} \left[ \mathrm{exp} \left(\frac{-r \beta_{\mathrm{d}}}{v_{\mathrm{exp}}}\right) - \mathrm{exp} \left(\frac{-r}{L_{\mathrm{p}}}\right) \right],
 \end{equation}

\noindent with $L_{\rm{p}}$ denoting the parent scale length, given by the ratio of the expansion velocity to the photodestruction rate of the parent species. For $L_{\rm{p}}$ = 0, equation (\ref{eq:haser_daughter_model}) reduces to equation (\ref{eq:haser_parent_model}).

We adopt expansion velocities of 0.7 km s$^{-1}$ for comet Lemmon and 1.0 km s$^{-1}$ for comet ISON, derived from the half-width at half maximum (HWHM) of the HCN lines, and kinetic gas temperatures of 55 K for comet Lemmon and 90 K for comet ISON \citep{Cordiner2014}. In Section \ref{subsec:line_ratio} we show how the line ratio of the HCN(4-3) and HCN(3-2) transitions constrain the kinetic temperature range to (20-110) K, and discuss how varying the kinetic temperature influences the molecular production rates we derive for each of the comets. We find that varying the temperature does not change the derived abundances significantly and therefore the temperatures of 55 K for Lemmon and 90 K for ISON are not critical parameters.

Photodestruction rates for HCN, CH$_3$OH and H$_2$CO are adopted from
\cite{Crovisier1994} (we assume that HNC has a similar photodestruction rate as HCN), H$_2$O from \cite{Budzien1994} and CS from \cite{Boissier2007}. Water production rates of (31.225$\pm$0.15)$\times 10^{28}$ s$^{-1}$ on 2013 May 11 and (17.68$\pm$0.26)$\times 10^{28}$ s$^{-1}$ on 2013 May 30 for Lemmon and (35.00$\pm$0.05)$\times 10^{28}$ s$^{-1}$ for ISON are deduced by \cite{Combi2014a, Combi2014} using the SOHO satellite.
 
As input, the LIME code takes molecular collision rates which we adopt from the Leiden Atomic and Molecular Database \cite[LAMDA;][]{Schoier2005}. The database holds collisional rates between H$_2$ and a number of the most abundant astronomical species. Since H$_2$O and not H$_2$ is the most important collisional partner in the inner part of cometary comae, we scale the LAMDA collision rates up with the hydrogen-to-water mass ratio of 9.0. To verify that this scaling does not bias our models, we vary the collisional scaling factor to investigate the effect. We find that for a collisional rate scaling factor higher than $\sim$5 our model outcome vary by only a few percent and that high scaling rate models converge to the outcome of a LTE model. The collisional rate scaling factor will be discussed further in Section \ref{subsec:line_ratio}. 
 
In the inner part of cometary comae the excitation of molecules is dominated by collisions. As the distance from the nucleus increases, densities drop and radiative processes, e.g., fluorescence through solar pumping, or collisions with electrons, become important. Here we focus only on the inner $\sim$3$\times$10$^3$ km of the coma. In this range the local density ratio of H$_2$O to electrons is very large and radiative processes negligible \cite[see][and references therein]{Bockelee-Morvan2004}; therefore we only consider collisions with H$_2$O in our model. To make our models computationally efficient we assume an outer cut-off of 5$\times$10$^3$ km.
 
To mimic the effect of the ALMA array, we run our model outputs though the tool "Simobserve" (part of the CASA package). By providing Simobserve with an antenna configuration file we sample our modelled sky brightness distribution with the same sampling function as that of the observations. We also simulate system noise and atmospheric effects in order to obtain a model as realistic and in accordance with observational effects as possible. After running Simobserve on all model outputs these are cleaned and imaged using the same parameters and routines as the observations.   

\subsection{Molecular production rates and parent scale lengths} \label{subsec:molecular_production_rates}

We derive molecular production rates for all detected species. To do this, we create a grid of models spanning a large range of molecular production rates and parent scale lengths. We then select the model that best reproduces the velocity integrated peak intensity and the specially integrated line flux in an aperture centred on each of the comets by minimising the $\chi^2$ values defined as the square of the difference between the observed and model flux divided by the square of the observational uncertainty. Our production rates are listed in Table \ref{tab:flux}.

We derive production rates of 6.7$\times10^{26}$ s$^{-1}$ for CS in comet ISON and (12.0-17.1)$\times10^{26}$ s$^{-1}$ and 17.0$\times10^{26}$ s$^{-1}$ for CH$_3$OH in comets Lemmon and ISON respectively. The CH$_3$OH production rates derived for comet Lemmon are based on four transitions. For the transitions CH$_3$OH(1$_{1,1}$-0$_{0,0}$), CH$_3$OH(7$_{1,6}$-6$_{1,5}$) and CH$_3$OH(7$_{2,6}$-6$_{2,4}$), which are observed on the same date, we report the weighted average of the production rate of the best-fit model of the individual transitions. The derived production rates have a standard deviation of 5.3 $\times$ 10$^{26}$ s$^{-1}$.

Using high-resolution spectroscopy \cite{DiSanti2016} derive production rates on ten pre-perihelion dates, including the dates of the observations presented here, of water and a number of trace molecules in ISON. They observe CH$_3$OH on 17 November and find a production rate of (16.0$\pm$3.0)$\times10^{26}$ s$^{-1}$ consistent with our value of (17.0$\pm$5.0)$\times10^{26}$ s$^{-1}$. On 19 November \cite{DiSanti2016} find that the production rate of CH$_3$OH has increased by a factor $\sim$2.5 to  (39.0$\pm$3.8)$\times 10^{26}$ s$^{-1}$. On similar dates (13-16 November), \cite{Agundez2014} derive a CH$_3$OH production rate of 43$\times 10^{26}$ s$^{-1}$ from observations carried out with the IRAM 30 m telescope. This range of CH$_3$OH production rate values clearly demonstrates the high variability of comets at decreasing heliocentric distances.

As mentioned in Section \ref{subsec:ISON}, we detect CS in the coma of comet ISON and derive a Q$_{\rm CS}$/Q$_{\rm H_2O}$ production rate ratio of 0.19$\%$. CS is not included in either of the studies by \cite{DiSanti2016} nor \cite{Agundez2014} but compared to the sample of four objects for which CS has been observed \cite[see Table 4 of][]{Mumma2011}, the CS-to-water production rate ratio we derive here is a little high, by $\sim$40$\%$, but consistent with CS being a product of CS$_2$ which is generally present in comets at a level of $\sim$(0.04-0.3) percent relative to water \citep{Cochran2015}.

Within errors, we derive production rates in agreement with those presented by \cite{Cordiner2014} for HCN and HNC in both comets. We derive a parent length scale of 1200 km for HNC in comet ISON, corresponding to a photodissociation rate of 8.3$\times10^{-4}$s$^{-1}$. Due to the low signal-to-noise of the HNC observation in comet Lemmon we are unable to distinguish between parent and daughter models, which both fit the data. Compared to the list of molecular production rates relative to water summarised by \cite{Bockelee-Morvan2004}, the rates we derive for HCN in each of the comets are consistent while our Q$_{\rm HNC}$/Q$_{\rm H_2O}$ values are lower by $\sim$40$\%$ and higher by $\sim$25$\%$ for comets Lemmon and ISON respectively. It should be noted that including the uncertainty in $L_{\mathrm{p}}$ for HNC in comet ISON results in a production rate range of (1.2-2.6)$\times$10$^{26}$ s$^{-1}$, within the errors of what is listed by \cite{Bockelee-Morvan1994}. The best-fit model for HCN in comet ISON has a non-zero parent length scale which is unexpected. Because of the size of the synthesised beam, structures in the coma of comet ISON which are smaller than $\sim$370 km will not be resolved. For $L_{\rm p}$=10 km, we derive a production rate of HCN in ISON of 3.7$\times$10$^{26}$ s$^{-1}$.
	
For H$_2$CO we derive production rates of 1.8$\times$10$^{26}$ s$^{-1}$ and 8.0$\times$10$^{26}$ s$^{-1}$ for Lemmon and ISON respectively. These values are both lower than what has been reported by \cite{Cordiner2014}. When modelling H$_2$CO, we only take into account ortho-H$_2$CO. Adopting an ortho-to-para ratio of 1 this implies that the production rates derived here only account for 50$\%$ of the H$_2$CO present in the comae of each of the comets and therefore the rates we report are lower limits. Taking the $L_{\mathrm{p}}$ uncertainties into account the range of production rates we derive for H$_2$CO are (1.3-2.4)$\times$10$^{26}$ s$^{-1}$ for Lemmon and (6.0-10.0)$\times$10$^{26}$ s$^{-1}$ for ISON. The parent length scales we derive for H$_2$CO are (1400$\pm$300) km for Lemmon and (250$\pm$50) km for ISON. These lengths correspond to photodissociation rates of 5$\times$10$^{-4}$ s$^{-1}$ and 4$\times$10$^{-3}$ s$^{-1}$ respectively, an order of magnitude higher that the photodissociation rate of CH$_3$OH \citep{Crovisier1994}. This mismatch between parent length scale and dissociation rate, combined with the methanol-to-water rate ratio of $\sim$1$\%$ for Lemmon and only $\sim$0.5$\%$ for ISON, makes CH$_3$OH an unlikely predecessor of H$_2$CO.

While the derived Q$_{\rm HCN}$/Q$_{\rm H_2O}$ ratios are fairly similar for comets Lemmon and ISON, other rate ratios vary. For instance, the Q$_{\rm CH_3OH}$/Q$_{\rm H_2O}$ rate ratio we derive for comet Lemmon is higher by about a factor two compared to that derived for comet ISON. This difference may however be due to the fact that we derive the production rate of CH$_3$OH in comet ISON based solely on the CH$_3$OH(1$_{1,1}$-0$_{0,0}$) transition whereas the rate derived for comet Lemmon uses multiple transitions. If we derive the production rate of CH$_3$OH in comet Lemmon only taking into account the CH$_3$OH(7$_{1,6}$-6$_{1,5}$) and CH$_3$OH(7$_{2,6}$-6$_{2,4}$) transitions, we derive a value which is 20$\%$ higher than the value we derive when we also include the CH$_3$OH(1$_{1,1}$-0$_{0,0}$) transition. On the other hand, the Q$_{\rm HNC}$/Q$_{\rm H_2O}$ and Q$_{\rm H_2CO}$/Q$_{\rm H_2O}$ rates are both higher in comet ISON compared to comet Lemmon. In particular, the Q$_{\rm HNC}$/Q$_{\rm H_2O}$ rate shows a difference of more than an order of magnitude between the two comets. As noted above, the uncertainty in the determination of the production rate of HNC in comet ISON, due to the uncertainty in $L_{\rm p}$, is not negligible. However, this uncertainty is insufficient to account for the large variation between the comets and thus underlines the importance of, and need for, comprehensive cometary studies to sample as many different bodies as possible in order to reveal the full range of cometary compositions. In their study \cite{Cordiner2017b} suggest that the observations of comet ISON may have coincided with the release of a clump of material rich in HNC (-precursor) molecules. Such an event may provide an explanation for the large Q$_{\rm HNC}$/Q$_{\rm H_2O}$ ration seen in comet ISON. It is also interesting to note that the production rate of H$_2$CO in comet ISON is high compared to the other detected species. This high rate may likewise be caused by an increase in the release of H$_2$CO-rich material, as the comet reaches perihelion shortly after the time of observation. In order to investigate similar events in the future, follow-up observations are essential.

\subsection{Formation scenarios of detected molecules}
\label{subsec:chemistry}

As discussed above, we identify HNC as a dughter species, formed in the expanding coma material. A gas-phase formation route of HNC is in agreement with the findings of \cite{Irvine1998} who determined the HNC/HCN ratio in the bright comet Hale-Bopp. They show that the HNC/HCN ratio varies with heliocentric distance in a way that is consistent with models of gas-phase chemical production of HNC but which cannot be explained if the HNC molecules are released directly from the cometary nucleus. Another possible formation route for HNC is via isomerisation of HCN through proton transfer reactions \citep{Rodgers1998}. However, in order for the HCN-HNC conversion to be efficient, hydrogen atoms in the cometary coma need to be suprathermal and the process is therefore insufficient to explain the HNC/HCN ratios observed in comets less active than Hale-Bopp. For these objects, the HNC/HCN ratio cannot be explained by neither ion-neutral coma chemistry nor isomerisation reactions \citep{Rodgers2001} and \cite{Rodgers2003} instead suggest that HNC is formed from degradation of some complex organic component, e.g., some variant of polymerized HCN, as also suggested by \cite{Cordiner2017b}.

From a sample of 14 moderately active comets at heliocentric distances spanning 0.1-1.5 AU, \cite{Lis2008} derive HNC/HCN ratios ranging from $\sim$0.03 to 0.3. They note that the HNC/HCN ratio is independent of the water production rate but dependent strongly on the heliocentric distance, with the largest ratios observed in objects with R$_{\mathrm{h}}$ < 0.8 AU. Here we derive a HNC/HCN ratio of $\sim$0.03 for comet Lemmon. This value is in the range of what is reported by \cite{Lis2008} and illustrates well the strong dependence of the HNC/HCN ratio on heliocentric distance (the heliocentric distance of comet Lemmon at the time of observations is $\sim$1.5AU). For comet ISON, we derive a HNC/HCN ratio of 0.45, consistent with ISONs small heliocentric distance and high HNC activity.

Observations of H$_2$CO in a number of cometary comae, both in-situ and from the ground, have likewise established that the radial profile of H$_2$CO cannot be explained by sublimation from the nucleus alone \cite[see][for an overview]{Cottin2008}. To account for the extended distribution of H$_2$CO a number of scenarios have been put forward. One possible formation route of H$_2$CO is through photodissociation of CH$_3$OH. This route is investigated by \cite{Cottin2004} who calculate the amount of CH$_3$OH required to account for the detected H$_2$CO in observations of comet 1P/Halley. Assuming H$_2$CO to be the main photoproduct of CH$_3$OH they find that CH$_3$OH would have to constitute 16$\%$ relative to water of the nucleus. Generally, CH$_3$OH is only found at levels up to $\sim$6.5$\%$ relative to water \citep{Cochran2015}, and we find ratios of $\sim$1$\%$ and $\sim$0.5$\%$ for Lemmon and ISON respectively. Moreover, they note that H$_2$CO is not the main dissociation product of CH$_3$OH but rather the CH$_3$O radial and therefore an abundance of CH$_3$OH of 16$\%$ is a lower limit. \cite{Cottin2004} therefore conclude that the extended source of H$_2$CO in comet Halley is inconsistent with a CH$_3$OH parent. As an alternative, \cite{Cottin2004} show that the distributed origin of H$_2$CO can be explained by photo- and/or thermal-degradation of refractory organic material in grains ejected from the cometary nucleus, with a simple addition-polymer of H$_2$CO as a likely candidate for the parent species. \cite{Cottin2004} are able to reproduce the observed distribution of H$_2$CO in comet Halley if the H$_2$CO-polymer is present in the grains at a mass fraction of a few percent. A similar explanation is given by \cite{Milam2006} who study the distribution of H$_2$CO in comets Hale-Bopp, Q4/NEAT and T7/LINEAR and likewise conclude that H$_2$CO must be coming from a source other than the nucleus. In contrast to \cite{Cottin2004} however, \cite{Milam2006} exclude a H$_2$CO-polymer as the most likely source for H$_2$CO, arguing that the required levels of such a polymer to efficiently account for the observed H$_2$CO are too high and that the possibility of forming such quantities of the polymers in interstellar space and the early solar system is low. Instead they propose that H$_2$CO is simply embedded in volatile grain matrices, entrained by solar radiation and released directly into the coma upon vaporization.

CS is thought to be a daughter species formed through photolysis of a parent molecule, such as CS$_2$, with a very short photodissociation lifetime of (10$^{2}$-10$^{3}$) s at 1 AU \cite[see][and references therein]{Feldman2004}. This short lifetime of the parent species indicates that CS is formed in the innermost part of the coma which is consistent with the observation of comet ISON were we find a spatial distribution of CS which is centrally peaked and coincide with the peak of the continuum emission. Due to the short lifetime of the CS parent molecule, high spatial resolution observations are needed to distinguish between parent and daughter formation scenarios. This is clearly illustrated when comparing the results presented here with those of previous studies, e.g., \cite{Snyder2001}, who use the BIMA array to characterise CS in comet Hale-Bopp, have identified CS as a parent species. In contrast, we derive a parent scale length of (200$\pm$50) km for the production of CS corresponding to a lifetime of the parent species of (1.5-2.5)$\times$10$^{2}$ s (assuming an outflow velocity of 1.0 km s$^{-1}$), consistent with the predictions of \cite{Feldman2004} and supporting a coma formation route for CS. The apparent discrepancy between the results can however be explained by the large difference in spatial resolution (the smallest BIMA beam diameter is more than twenty times that of the synthesised ALMA beam). The new results therefore clearly demonstrate the superior resolving power of ALMA compared to previous interferometric arrays. However, it should be noted that with a synthesised beam of 0$\overset{\second}{.}$59$\times$0$\overset{\second}{.}$39 for the CS observations, corresponding to $\sim(375\times250)$ km at the distance of comet ISON, a parent scale length of 200 km is on the limit of what can be resolved. To unambiguously determine the origin of CS, higher spatial resolution observations are therefore essential. 

\subsection{Integrated intensity maps} \label{subsec:integrated_intenisity_maps}
In the case of comet Lemmon our models reproduce total integrated fluxes within $\sim$20$\%$ for HCN and $\sim$60$\%$ for HNC. For CH$_3$OH the total integrated flux of the CH$_3$OH(1$_{1,1}$-0$_{0,0}$) and CH$_3$OH(5$_{2,3}$-4$_{1,3}$) transition are overestimated by $\sim$10$\%$ and $\sim$30$\%$ respectively whereas the modelled CH$_3$OH(7$_{1,6}$-6$_{1,5}$) and CH$_3$OH(7$_{2,6}$-6$_{2,4}$) transitions are under-predicted with about 15$\%$ with respect to the observed values. The fact that our model under-predicts the total intensity suggests that we are not treating extended emission in a correct way. This is not surprising since we use a single parameter to express the outflow velocity of molecules. Such a single-parameter approach may very well be too much of a simplification, as indicated in the study of comet 17P/Holmes by \cite{Qi2015}, where two distinct components in the molecular emission are identified. Adopting a lower outflow velocity will result in a higher abundance of material in the outer coma relative to a high outflow-velocity scenario and may provide a better fit to the data. However, introducing a variable temperature and velocity model, such as the one presented by \cite{Friedel2005}, is beyond the scope of this paper. In addition, as shown by \citet{Cordiner2017b} for comet ISON, temporal variations on time scales $<20$ min affect the production of the cometary species. This means that our peak intensity and integrated intensity sample different time averages which cannot necessarily be matched by a single time averaged production rate. In the case of comet ISON total integrated fluxes are reproduced within $\sim$5$\%$ of all observations.

\subsection{HCN(4-3)/(3-2) line ratio} \label{subsec:line_ratio}
The uncertainty in the production rates we derive arise from uncertainties in excitation caused by the unknown hydrogen-to-water collisional rate scaling and coma temperature. As a means to constrain these uncertainties the ratio between lines can be of help. In the data presented here multiple HCN transitions are observed, albeit not on the same date. With the combination of observations in Band 6 and 7 of comet Lemmon we can study the line ration of HCN(4-3) to HCN(3-2). Using the velocity integrated peak intensity of each of the transitions we derive a line ratio for HCN(4-3)/(3-2) of 2.40$\pm$0.34. 

Ideally we want to observe multiple transitions of a single molecule (almost) simultaneously ensuring that transitions are observed under similar physical conditions. Unfortunately this is not the case for our observations where the HCN(4-3) and HCN(3-2) transitions are observed three weeks apart. Because of this, excitation conditions are not directly comparable i.e. water production rates vary by a factor $\sim$2 \citep{Combi2014a}. 

As a workaround, we scale the HCN production rate of one set of observations to the other under the following assumptions; firstly, that the coma temperature and collision rate scaling are the same on both dates; secondly, that the water production rates on the individual dates are known, and thirdly, that the HCN/H$_2$O ratio is constant and does not change over time. Since HCN-variations are small between objects \cite[see, e.g.,][]{Bockelee-Morvan2004,Mumma2011,Cochran2015}, it is reasonable to assume that HCN is constant within a single source. These assumptions allow us to calculate the HCN(4-3) and HCN(3-2) lines for their respective dates (assuming the same, but unknown, HCN/H$_2$O ratio), and explore the effects of the adopted kinetic temperature and collisional rate scaling on the inferred Q$_{\rm HCN}$/Q$_{\rm H_2O}$ ratio. 

To investigate the extent to which the detection of a set of molecular transitions, and the ratio between them, can constrain the molecular production rate and kinetic temperature of a comet better than an individual transition, we construct a grid of models and calculate the HCN(4-3)/(3-2) line ratio for each grid cell varying the HCN production rate and kinetic temperature. We vary the temperature between 5 and 150 K and the Q$_{\rm HCN}$/Q$_{\rm H_2O}$ ratio between 10$^{-4}$ and 1. 

\begin{figure}[]
	\includegraphics[width=0.55\textwidth]{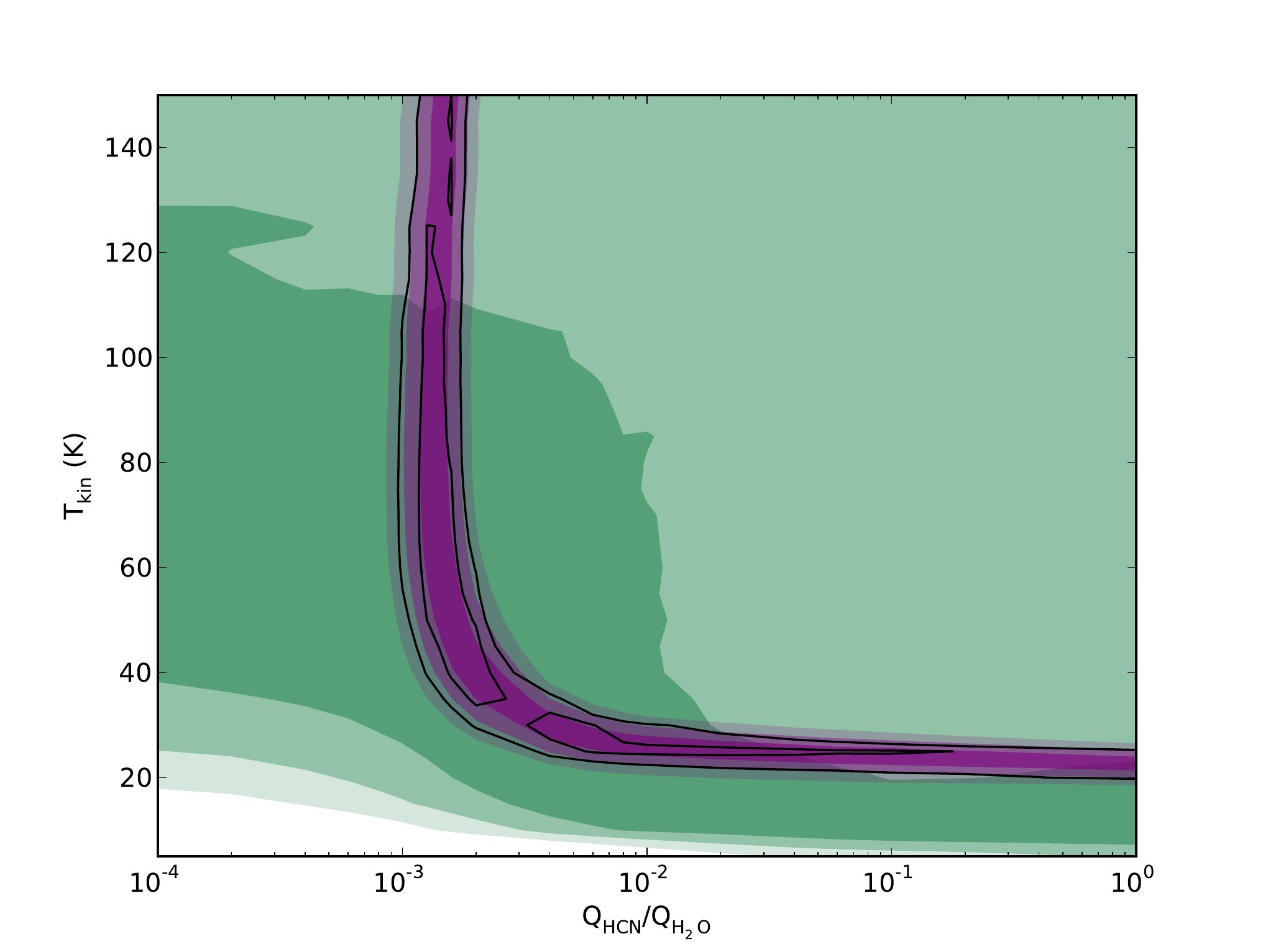}
	\caption[HCN(4-3)/(3-2) line ratio as function of pruduction rate and kinetic temperature]{Model grid of HCN(4-3)/(3-2) line ratio as function of production rate and kinetic temperature. Green and purple colour scales represent the 1, 2 and 3 $\sigma$ ranges of the HCN(4-3)/(3-2) line ratio and HCN(4-3) velocity integrated peak intensity respectively. The black solid lines indicate the reduced $\chi^2$ contours equal to 1 and 3.}
	\label{fig:model_constant_coll}
\end{figure}

Figure \ref{fig:model_constant_coll} shows the model grid with the HCN(4-3)/(3-2) line ration in green and the HCN(4-3) velocity integrated peak intensity in purple. The green and purple colour gradients indicate the 1, 2 and 3 $\sigma$ ranges. We adopt a conservative 10$\%$ error on the absolute flux calibration which we add to the RMS noise in quadrature. In the same plot, the solid black lines show the reduced $\chi^2$ contours equal to 1 and 3. 

Our models show that the Q$_{\rm HCN}$/Q$_{\rm H_2O}$ ratio is very well constrained to (0.5-3.0)$\times$10$^{-3}$ for kinetic temperatures above 40 K. The HCN(4-3)/(3-2) line ratio excludes temperatures above 110 K at 1$\sigma$; higher temperatures are allowed at lower significance. 

If the temperature drops below 30 K the Q$_{\rm HCN}$/Q$_{\rm H_2O}$ rate increases rapidly. This means that very cold regions ($<$30 K) are not likely to contribute to the observed emission, as they would require Q$_{\rm HCN}$/Q$_{\rm H_2O}$ ratios much larger than commonly found in comets. It also shows that the observations are not very sensitive to such cold material in the coma. Temperatures below 20 K can be ruled out entirely for the HCN line emitting vapor. Together, this shows that the adopted temperature of 55 K is not a critical parameter in determining Q$_{\rm HCN}$/Q$_{\rm H_2O}$ except if we consider very cold regions in the coma.

\begin{figure}[]
\includegraphics[width=0.55\textwidth]{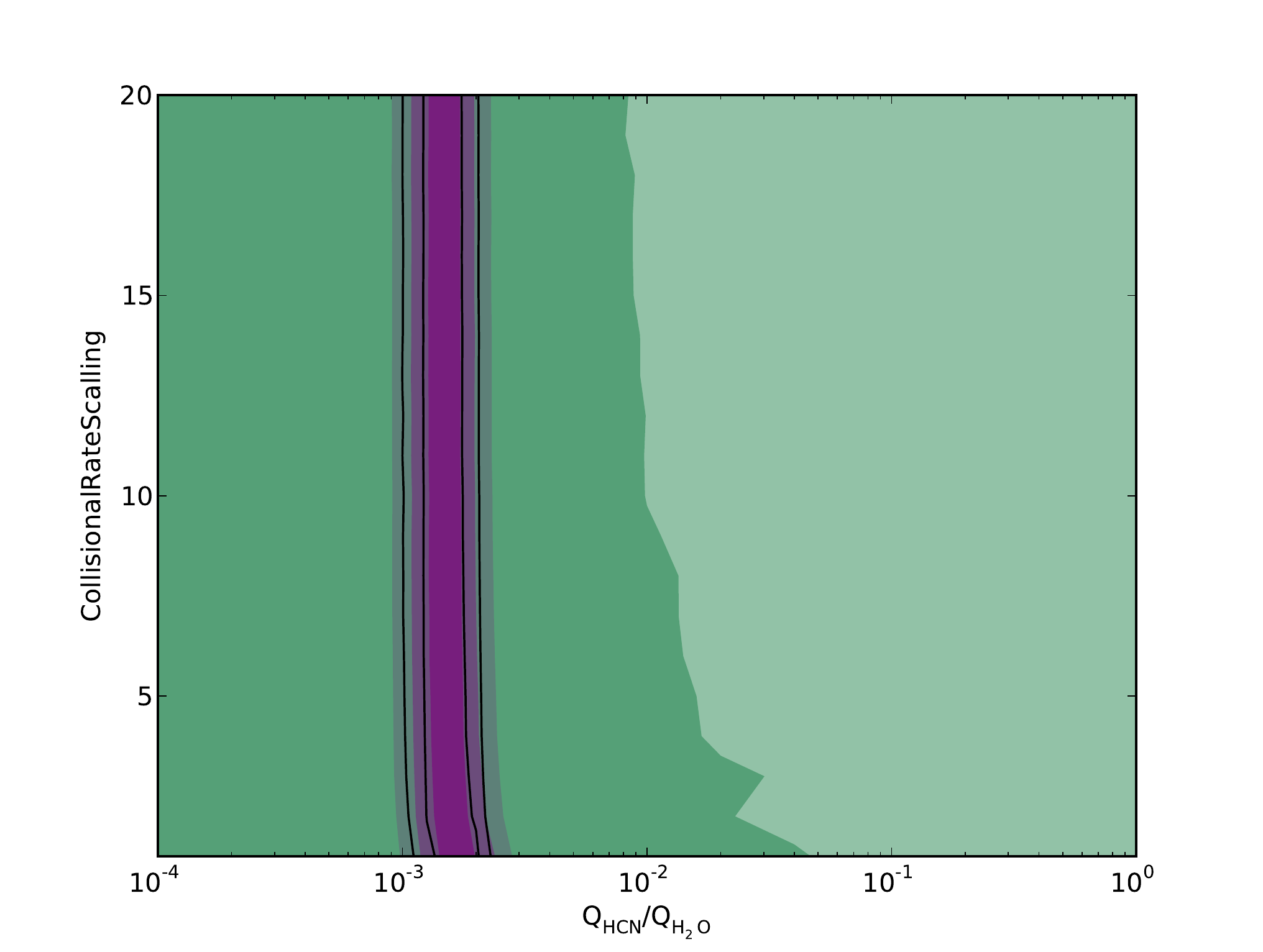}
\caption[HCN(4-3)/(3-2) line ratio as function of pruduction rate and collisional rate scaling factor]{Model grid of HCN(4-3)/(3-2) line ratio as function of production rate and collisional rate scaling factor. Green and purple colour scales represent the 1,2 and 3 $\sigma$ ranges of the HCN(4-3)/(3-2) line ratio and HCN(4-3) velocity integrated peak intensity respectively. The black solid lines indicate the reduced $\chi^2$ contours equal to 1 and 3.}
\label{fig:model_constant_temp}
\end{figure}

To investigate the effect of the collisional rate scaling on Q$_{\rm HCN}$/Q$_{\rm H_2O}$, we construct a model grid varying the collisional rate scaling factor between 1 and 20. Figure \ref{fig:model_constant_temp} shows the model gird. Colours and lines are the same as in Fig. \ref{fig:model_constant_coll}. It is clear that the Q$_{\rm HCN}$/Q$_{\rm H_2O}$ ratio is not sensitive to the collisional rate scaling for scaling factors higher than $\sim$5. This shows that our adopted collisional rate scaling factor of 9.0 does not play a vital role when determining Q$_{\rm HCN}$/Q$_{\rm H_2O}$.  

From Figs. \ref{fig:model_constant_coll} and \ref{fig:model_constant_temp} it is evident that the Q$_{\rm HCN}$/Q$_{\rm H_2O}$ ratio is well constrained as long as temperatures below 30 K can be excluded and the collisional rate scaling factor is kept above 5.  

Because we do not have multiple transitions of other species, we use the constraints on kinetic temperature and collisional rate scaling factor derived from the HCN(4-3)/HCN(3-2) line ratio to investigate how the derived molecular production rates of other species are effected. To do this, we run a set of four models representing each of the extreme cases; high temperature ($T$ = 110 K (Lemmon) and $T$ = 180 K (ISON)) and high collisional rate scaling (scaling rate factor = 20), high temperature and low collision rate scaling (scaling rate factor = 1), low temperature ($T$ = 20 K) and high collisional rate scaling and low temperature and low collisional rate scaling. We run models for CS, HCN and CH$_3$OH in comet ISON and CH$_3$OH in comet Lemmon. We do not run models for the distributed sources, HNC and H$_2$CO, because the uncertainty on the production rates derived here is dominated by uncertainties in $L_{\rm p}$. 

For collisional scaling rates less than 5, the molecular production rate vary within 3$\%$ for CS, 9$\%$ for HCN and 26$\%$ for CH$_3$OH. For collisional scaling rates higher than 5 the molecular production rate become independent of collisional rate scaling. This is reassuring for the robustness of our model but also illustrates that even if the molecular production rate of a specific species can be very well constrained, no constraint can be set on the collisional cross section of water with respect to this species. Thus the precise determination of collisional cross sections of water with respect to other species presents a challenge for future laboratory works. Until such collisional cross section are available, ALMA provides a particularly beneficial tool since its small beam allows us to probe only the innermost part of the cometary coma where molecular densities are high enough to ensure that excitations are close to LTE, i.e., the exact values of the collisional rates are no longer significant.

In contrast to the case of the collisional rate scaling factor, we do see a dependence of the derived production rate on kinetic temperature. In the case of CS and HCN, the high temperature case returns production rates within $\sim$40$\%$ and $\sim$50$\%$ of those derived using the temperatures of 55 K and 90 K for comets Lemmon and ISON respectively (with the derived molecular production rates ranging (1.7-3.0)$\times$10$^{26}$ s$^{-1}$ for HCN in comet Lemmon and (6.0-9.5)$\times$10$^{26}$ s$^{-1}$ for CS, and (3.5-5.9)$\times$10$^{26}$ s$^{-1}$ for HCN in comet ISON). In the low temperature cases, production rates need to be unreasonably high, i.e., orders of magnitude higher than what is commonly observed in comets, in order to reproduce the observations. It is important to note that if very low-temperature regions exist in the coma, e.g., as a result of adiabatic expansion, molecular line observations are essentially `blind' to them. As shown by \citet{Cordiner2017a} for comet C/2012 K1 (PanSTARRS), such very cold regions may exist in the cometary coma and it is therefore important for future studies to characterize the coma temperature profile in order to derive reliable molecular production rates. In the case of CH$_3$OH, the high and low temperature models result in a range of derived production rates of (4.1-36.3)$\times$10$^{26}$ s$^{-1}$ for comet Lemmon and (2.4-45.0)$\times$10$^{26}$ s$^{-1}$ for comet ISON.

From our analysis, it is clear that while the collisional rate scaling factor only plays a minor role when determining molecular production rates, and has no effect on the model outcome when they are higher than 5, an accurate estimate of the kinetic gas temperature is important. It is also clear that multiple transitions, observed simultaneously, can help constrain molecular production rates considerably. 

\section{Conclusion} \label{sec:conclusion}
In this paper we analyse archival ALMA observations of volatile species in the comae of comets Lemmon and ISON. We report the first ALMA detection of CS and in the coma of comet ISON as well as several CH$_3$OH transitions and the HCN(4-3)/HCN(3-2) line ratio in comet Lemmon. In addition we confirm the detection of HCN, HNC and H$_2$CO in each comet, as reported by \cite{Cordiner2014}.

We derive a parent scale length of (200$\pm$50) km for CS. A structure of this size is similar to that of the ALMA synthesised beam and therefore supports the theory of CS being a daughter rather than parent species.

We have mapped the spatial distribution of each molecule and find centrally peaked, symmetric distributions for HCN and CH$_3$OH, indicative of parent species, i.e., species sublimated directly from the cometary nucleus. In contrast we see distributed origins for HNC and H$_2$CO in both comets consistent with these species being either the result of gas-phase chemistry in the coma or transported away from the nucleus by some refractory compound before being evaporated.

To model the individual line transition intensities we use the 3D radiative transfer code LIME, assuming a Haser profile with constant outflow velocity to represent the density of molecules in the comae. Based on these models we derive molecular production rates for each of the detected species. Assuming a kinetic gas temperature of 55 K for comet Lemmon and 90 K for comet ISON we derive production rates of 6.7$\times$10$^{26}$ s$^{-1}$ for CS, (2.0-4.0)$\times$10$^{26}$ s$^{-1}$ for HCN, (0.06-1.8)$\times$10$^{26}$ s$^{-1}$ for HNC, (1.8-8.0)$\times$10$^{26}$ s$^{-1}$ for H$_2$CO and (12.0-17.1)$\times$10$^{26}$ s$^{-1}$ for CH$_3$OH. Our derived production rates are consistent with production rates relative to water in the literature \citep[e.g.][]{Mumma2011,Bockelee-Morvan2004}.

Because collisional cross sections of water with respect to the species detected here are unknown, we scale the H$_2$ collisional rates up with the hydrogen-to-water mass ratio of 9.0. We investigate the effect of this scaling on the derived production rates and conclude that the scaling factor is not a critical parameter in our model. The independence of the derived production rate with respect to the collisional rate scaling is a consequence of the small ALMA beam which allows us to only sample the inner and high density region of the coma where excitations are close to LTE and the exact value of the water collisional cross sections therefore become less important. 

Under the assumption of a constant temperature and Q$_{\rm HCN}$/Q$_{\rm H_2O}$ ratio, simultaneous modelling of the HCN(3-2) line of 2013 May 11 and the HCN(4-3) line of 2013 June 1, shows that the line ratio of HCN(4-3)/HCN(3-2) provides good constraints on the kinetic gas temperature and molecular production rate. The HCN(4-3)/HCN(3-2) line ratio excludes temperatures above 110 K at 1$\sigma$, while temperatures below 30 K would require unusually large Q$_{\rm HCN}$/Q$_{\rm H_2O}$ ratios in order to reproduce the observations. Observations are not sensitive to material below 20 K. With the constraints on temperature derived from the HCN(4-3)/HCN(3-2) line ratio, we investigate the effect of this range on the derived production rate of other species. In both comets Lemmon and ISON the production rate of HCN is reproduced within 50$\%$ in the high-temperature scenario, while the low-temperature scenario can be excluded completely. For CS this number is 40$\%$. In the case of CH$_3$OH, we find a variation in the derived production rate from the low to high temperature case corresponding to an order of magnitude. This illustrates how the constraints on temperature derived from multi-transition observations of cometary volatiles can improve the accuracy of inferred production rates.

\begin{acknowledgements} 
The authors would like to thank Dr. Martin Cordiner for valuable discussions and the anonymous referee for constructive comments that significantly improved our manuscript. We acknowledge data reduction support from Allegro, the European ALMA Regional Center node in the Netherlands, and Markus Schmalzl in particular for expert advice. This paper makes use of the following ALMA data: ADS/JAO.ALMA\#2012.A.00020.S, \#2012.A.00033.S and \#2011.0.00012.SV. ALMA is a partnership of ESO (representing its member states), NSF (USA) and NINS (Japan), together with NRC (Canada) and NSC and ASIAA (Taiwan) and KASI (Republic of Korea), in cooperation with the Republic of Chile. The Joint ALMA Observatory is operated by ESO, AUI/NRAO and NAOJ
\end{acknowledgements}

\bibliographystyle{aa} 
\bibliography{Lemmon_and_ISON} 

\end{document}